\documentclass[runningheads,psfig]{svmult}

\usepackage{makeidx}   
\usepackage{graphicx}  
\usepackage{subeqnar}  
\usepackage{multicol}  
\usepackage{physprbb}  
\makeindex             

\newcommand{\HI}{H\thinspace\protect\footnotesize
I\protect\normalsize}

\newcommand{\B}{{$B$}}

\newcommand{\II}{{$I_c$}}
\newcommand{\J}{{$J$}}
\newcommand{\HH}{{$H$}}
\newcommand{\K}{{$K_s$}}
\newcommand{\tfr}{Tully\,--\,Fisher relation}
\newcommand{\kms}{\,km\,s$^{-1}$}
\newcommand{\etal}{{et~al.\, }}
\newcommand{\cf}{{cf.\, }}
\newcommand{\eg}{{e.g.\, }}         
\newcommand{\ie}{{i.e.\, }}         

\newcommand{\GA}{\,\raisebox{-0.4ex}{$\stackrel{>}{\scriptstyle\sim}$}\,}
\newcommand{\LA}{\,\raisebox{-0.4ex}{$\stackrel{<}{\scriptstyle\sim}$}\,}

\def\deg{\hbox{$^\circ$}}
\def\arcmin{\hbox{$^\prime$}}
\def\arcsec{\hbox{$^{\prime\prime}$}}
\def\micron{\hbox{$\mu$m}}

\def\fm{\hbox{$.\!\!^{\rm m}$}}

\def\fdg{\hbox{$.\!\!^\circ$}}
\def\farcm{\hbox{$.\mkern-4mu^\prime$}}
\def\farcs{\hbox{$.\!\!^{\prime\prime}$}}

\begin{document}
\title*{Galaxies Behind the Milky Way \protect\newline 
and the Great Attractor}
\toctitle{Galaxies Behind the Milky Way and the Great Attractor}
%
%
\titlerunning{Galaxies Behind the Milky Way}
%
\author{Ren\'ee C. Kraan-Korteweg}
\authorrunning{Ren\'ee C. Kraan-Korteweg}
%
%
\institute{Departamento de Astronom\'{\i}a, Universidad de Guanajuato,
Apartado Postal 144, 36000 Guanajuato GTO, Mexico}

\maketitle              

\begin{abstract}
Dust and stars in the plane of the Milky Way create a "Zone of
Avoidance" in the extragalactic sky. Galaxies are distributed in
gigantic labyrinth formations, filaments and great walls with
occasional dense clusters. They can be traced all over the sky, except
where the dust within our own galaxy becomes too thick -- leaving about
25\% of the extragalactic sky unaccounted for. Our Galaxy is a natural
barrier which constrains the studies of large-scale structures in the
Universe, the peculiar motion of our Local Group of galaxies and other
streaming motions (cosmic flows) which are important for understanding
formation processes in the Early Universe and for cosmological models.
 
Only in recent years have astronomers developed the techniques to peer
through the disk and uncover the galaxy distribution in the Zone of
Avoidance. I present the various observational multi-wavelength
procedures (optical, far infrared, near infrared, radio and X-ray)
that are currently being pursued to map the galaxy distribution behind
our Milky Way, including a discussion of the (different) limitations
and selection effects of these (partly) complementary approaches.  The
newly unveiled large-scale structures are discussed and compared
to predictions from theoretical reconstructions of the mass density
field. Particular emphasis is given to discoveries in the Great
Attractor region -- a from streaming motions predicted huge
overdensity centered behind the Galactic Plane. The recently unveiled
massive rich cluster A3627 seems to constitute the previously
unidentified core of the Great Attractor.
\end{abstract}

\section{The Zone of Avoidance}

A first reference to the Zone of Avoidance (ZOA), or the ``Zone of few
Nebulae'' was made in 1878 by Proctor \cite{Pro78}, based on the distribution of
nebulae in the ``General Catalogue of Nebulae'' by Sir John Herschel
\cite{Her64}. 
This zone becomes considerably more prominent in the
distribution of nebulae presented by Charlier \cite{Char22} 
using data from
the ``New General Catalogue'' by Dreyer \cite{Dre88,Dre95}. 
These data also reveal
first indications of large-scale structure: the nebulae display a very
clumpy distribution. Currently well-known galaxy clusters such as 
Virgo, Fornax, Perseus, Pisces and Coma are easily recognizable even
though Dreyer's catalog contains both Galactic and extragalactic objects
as it was not known then that the majority of the nebulae actually
are external stellar systems similar to the Milky Way. Even more
obvious in this distribution, though, is the absence of galaxies around
the Galactic Equator. As extinction was poorly known at that time, no
connection was made between the Milky Way and the ``Zone of few
Nebulae''.

A first definition of the ZOA was proposed by Shapley  \cite{Sha61}, 
as the
region delimited by ``the isopleth of five galaxies per square degree
from the Lick and Harvard surveys'' (compared to a mean of 54
gal./sq.deg. found in unobscured regions by Shane \& Wirtanen \cite{Sha67}).
This ``Zone of Avoidance'' used to be ``avoided'' by
astronomers interested in the extragalactic sky because of the
inherent difficulties in analyzing the few obscured galaxies known
there.

Merging data from more recent galaxy catalogs, \ie the Uppsala General
Catalog UGC \cite{Nil73} 
for the north ($\delta \ge -2\fdg5$), the
ESO Uppsala Catalog \cite{Lau82} 
for the south ($\delta \le
-17\fdg5$), and the Morphological Catalog of Galaxies MCG \cite{Vor63}
for the strip inbetween
($-17\fdg5 < \delta < -2\fdg5$), a whole-sky galaxy catalog can be
defined. To homogenize the data determined by different groups
from different survey material, the following adjustments have to be
applied to the diameters: ${D = 1.15 \cdot D_{\rm UGC}, D = 0.96 \cdot
D_{\rm ESO}}$ and ${D = 1.29 \cdot D_{\rm MCG}}$  \cite{Fou85}.
According to Hudson \& Lynden-Bell \cite{Hud91} 
this ``whole-sky''
catalog then is complete for galaxies larger than ${D} =1\farcm3$.

\begin{figure}[ht]
\begin{center}
\includegraphics[width=12cm]{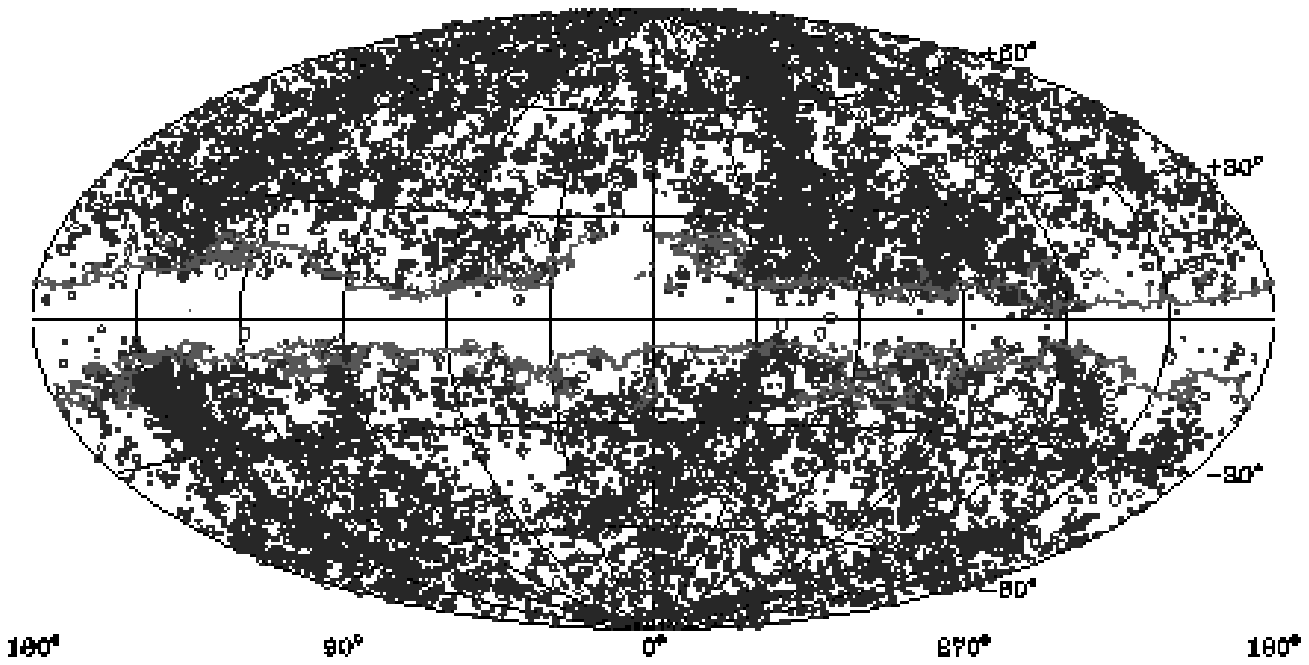}
\caption
{Aitoff equal-area projection in Galactic coordinates of galaxies with
${D}\ge1\farcm$3. The galaxies are diameter-coded: small circles
represent galaxies with $1{\farcm}3 \le {D} < 2\arcmin$, larger
circles $2\arcmin \le {D} < 3\arcmin$, and big circles ${D}
\ge 3\arcmin$. The contour marks absorption in the blue of ${A_B}
= 1\fm0$ as determined from the Schlegel \etal [13]
dust extinction
maps.  The displayed contour surrounds the area where the galaxy
distribution becomes incomplete (the ZOA) remarkably well}
\label{ait}
\end{center}
\end{figure}

The distribution of these galaxies is displayed in Galactic
coordinates in Fig.~\ref{ait} in an equal-area Aitoff projection
centered on the Galactic Bulge ($\ell = 0\deg, b = 0\deg$). The
galaxies are diameter-coded, so that structures relevant for the
dynamics in the local Universe stand out accordingly. Most conspicuous
in this distribution is, however, the very broad, nearly empty band of
about 20$\deg$. Why this Zone of Avoidance?  Optical galaxy catalogs are
limited to the largest galaxies. They therefore become increasingly
incomplete close to the Galactic Equator where the dust thickens. This
diminishes the light emission of the galaxies and reduces their
visible extent. Such obscured galaxies are not included in diameter- or
magnitude-limited catalogs because they appear small and faint -- even
though they might be intrinsically large and bright. A further
complication is 
the growing number of foreground stars close to the Galactic Plane
(GP) which fully or partially block the view of galaxy images.

Comparing this ``band of few galaxies'' with the currently available
dust extinction maps of the DIRBE experiment \cite{Sch98},
we can see that the ZOA -- the area where the galaxy
counts become severely incomplete -- is described almost perfectly by
the absorption contour in the blue ${A_B}$ of $1\fm0$ (where ${A_B}$
is 4.14 times the extinction $E(B-V)$ \cite{Car89}).
This contour matches the ZOA defined by Shapley \cite{Sha61} 
closely.

\subsection {Constraints due to the Milky Way}
Why is the distribution of galaxies behind the Milky Way important, and
why is it not sufficient to study galaxies and their large-scale
distribution away from the foreground ``pollution'' of the Milky Way?  

In the last 20 years, enormous effort and observation time has been
devoted to map the galaxy distribution in space.  It was found that
galaxies are located predominantly in clusters, sheets and filaments,
leaving large areas devoid of luminous matter (see \cite{Fai98a}
for a detailed observational description of ``Large-Scale Structures in
the Universe'').

Our Galaxy is part of the Local Group (LG) of galaxies, a small,
gravitationally bound group of galaxies consisting of a few bright
spiral galaxies and about 2 dozen dwarf galaxies. Our LG lies in the
outskirts of the Local Supercluster, a flattened structure of about
30~Mpc, centered on the Virgo galaxy cluster with a few thousand
galaxies (including its numerous dwarfs). Many such superclusters have
meanwhile been charted. The nearby ones can actually be identified in
the 2-dimensional galaxy distribution of Fig.~\ref{ait}: the Local
Supercluster is visible as a great circle (the Supergalactic Plane)
centered on the Virgo cluster at $\ell=284\deg, b=74\deg$, the
Perseus-Pisces supercluster which bends into the ZOA at $\ell=95\deg$
and $\ell=165\deg$, and the general galaxy overdensity in the Great
Attractor (GA) region ($280 \LA \ell \LA 360\deg, |b| \LA 30\deg$).
Most of these superclusters and wall-like structures have massive
clusters at their centers.

The lack of data in the ZOA severely constrains the studies of these
structures in the nearby Universe, the origin of the peculiar velocity
of the Local Group, and other streaming motions.  Such studies are
dependent on an accurate description of the whole sky distribution of
galaxies, as described in the following sections.

\subsubsection{Peculiar Motion of the Local Group of Galaxies.}\label{vpec}

The Cosmic Microwave Background radiation (CMB) of $2.7\deg$~K -- the
relic radiation of the hot early Universe -- shows a dipole of about
0.1\%. This dipole is explained by a peculiar motion of the LG on top
of the uniform Hubble expansion of 630~\kms\ towards the Galactic
coordinates $\ell = 268\deg, b = 27\deg$ \cite{Kog93} 
induced by
the gravitational attraction of the irregular mass distribution in the
nearby Universe (see Fig.~\ref{ait}). Part of this motion can be
explained by the acceleration of the LG towards Virgo, the center of
the Local Supercluster ($\sim 220$~\kms\ towards $\ell = 284\deg, b =
75\deg$). The remaining component of $\sim 495$~\kms\ towards $\ell =
274\deg, b = 12\deg$ \cite{San84,Sha84} 
hence must
arise from other mass concentrations and/or voids in the nearby
Universe.  The determination of the peculiar motion on the LG, \ie its
net gravity field, requires whole-sky coverage. Here, the lack of data
in about 25\% of the optical extragalactic sky is a severe handicap.

Various dipole determinations have assumed a uniformly filled ZOA or
have used cloning methods which transplant the fairly well-mapped
adjacent regions into the ZOA. Both procedures are unsatisfactory,
because inhomogeneous data coverage will introduce non-existing flow
fields. The derived results on the apex of the LG motion, as well as
the distance at which convergence is attained, still are
controversial. Kolatt \etal \cite{Kol95},  
for instance, have shown that the
mass distribution within the inner $\pm20\deg$ of the ZOA -- as
derived from theoretical reconstructions of the density field (see
Sect.~\ref{recon}) -- is crucial to the derivation of the
gravitational acceleration of the LG: the direction of the motion
measured within a volume of 6000~\kms\ will change by $31\deg$ when
the (reconstructed) mass within the ZOA is included. Care should
therefore be taken on how to extrapolate the galaxy density field
across the ZOA. Obviously, a reliable consensus on the galaxy
distribution in the ZOA is important to minimize these uncertainties.

\subsubsection{Nearby Galaxies.}

In this context, not only the identification of unknown and suspected
clusters, filaments and voids are relevant, but also the detection of
nearby smaller entities. The peculiar velocity of the LG, ${\vec
v_p}$, is proportional to the net gravity field $\vec{G}$, which can
be determined by summing up the masses $\cal M_{\rm i}$ of the
individual galaxies at their distances $\vec{r_i}$:
$${\vec v_p} \propto f({\vec G}) \propto {\Omega_0^{0.6} \over b} \sum
{{\cal M}_{i} \over r_{i}^2} \, {\hat {\bf r_i}},$$ where
$\Omega_0$ is the density parameter and $b$ the bias parameter.  The
gravity field as well as the light flux of a galaxy decreases with
$r^{-2}$. The direction and amplitude of the peculiar velocity
therefore is directly related to the sum of the {\sl apparent
magnitudes} of the galaxies in the sky through
$$\vec{v_p} \propto \sum_{\rm i} 10^{-0.4{\rm m}} \, {\hat {\bf
r_i}},$$ 
for a constant mass-to-light ratio. This has important implications
and suggests, for instance, that the galaxy Cen A with an
absorption-corrected magnitude of $B^o = 6\fm1$ exerts a stronger
luminosity-indicated gravitational attraction on the Local Group than
the whole Virgo cluster. However, in this context, the question
whether the mass-to-light ratio is constant, \ie no biasing occurs, is
doubtful, a problem inherent to all cumulative dipole
determinations. These calculations also predict that the 8 apparently
brightest galaxies -- which are all nearby ($v < 300$~\kms) -- are
responsible for 20\% of the total dipole as determined from optically
known galaxies within $v \LA 6000$~\kms. Hence, a major part of the
peculiar motion of the LG is generated by a few average, but nearby
galaxies.

In this sense, the detection of other nearby galaxies hidden by the
obscuration of the Galaxy can be as important as the detection of
entire clusters at larger distances. The expectation of finding
additional nearby galaxies in the ZOA is not unrealistic. Six of the
nine apparently brightest galaxies are located in the ZOA: IC342,
Maffei 1 and 2, NGC4945, CenA and the recently discovered galaxy
Dwingeloo 1 (see Sect.~\ref{dogs}).  Moreover, the presence of an
unknown Andromeda-like galaxy behind the Milky Way would have
implications for the internal dynamics of the LG, the mass
determination of the LG, and the present density of the Universe from
timing arguments \cite{Pee94}. 

\subsubsection{Cosmic Flow Fields such as in the Great Attractor Region.} 
\label{flow}

Density enhancements locally decelerate the uniform expansion field, as
has been observed within our own Local Supercluster. Vice versa,
systematic streaming motions over and above the uniform expansion
field usually indicate mass overdensities (accelerations) or
voids (decelerations).  Knowing (a) the observed recessional velocity
$v_{\rm obs}$ of a galaxy through its redshift $z$
$$ v_{\rm obs} = c z = 
c \, {{\lambda (t) \, - \, \lambda_{0}}\over{\lambda_{0}}},$$
where $\lambda_{0}$ is the rest wavelength, and $\lambda (t)$ is the
observed wavelength, and (b) a redshift-independent distance estimate
$r$, the peculiar motion of a galaxy ${\vec v_p}$ due to the
underlying mass density field can be determined:
$${\vec v_p} = {\vec v_{\rm obs}} - {\vec v_{\rm Hub}},$$ 
where $v_{\rm Hub}$ is the recession velocity a galaxy would
have in an unperturbed expansion field ($v_{\rm Hub} = H_0 \cdot
r$). In this manner, the mass density field can be determined
independent of the galaxy distribution and/or an assumption on the
mass-to-light ratio.

Based on these considerations, Dressler \etal \cite{Dre87} 
identified a
systematic infall pattern from peculiar velocities of about 400
elliptical galaxies which was interpreted as being due to a
hypothetical Great Attractor with a mass of $\sim~5\times 10^{16}{\cal
M}_\odot$, at a position in redshift space of $(\ell,b,v) =
(307\deg,9\deg,\sim 4400$~\kms) \cite{Lyn88}. 
A more
recent study by Kolatt \etal \cite{Kol95}, 
based on a larger data set
(elliptical {\sl and} spiral galaxies) and the potential
reconstruction method POTENT (see Sect.~\ref{recon} and
Fig.~\ref{kolatt}) place the center of the GA right behind the Milky
Way. Recent consensus is that the GA is an extended region ($\sim
40\deg {\rm x} 40\deg$) of moderately enhanced galaxy density centered
behind the Galactic Plane.  Although there is a considerable excess of
optical galaxies and IRAS-selected galaxies in this region (see
Fig.~\ref{ait} and Fig.~\ref{BTP}), no dominant cluster or central
peak can been seen. However, a major part of the GA is hidden by the Milky
Way.

\subsubsection{Connectivity of Superclusters Across the ZOA.}\label{LSS}

Various large-scale structures are `bisected' by the Milky Way.  What
is their true extent? These large-scale structures, their sizes, and
the distribution of the various galaxy types within these structures,
carry information on the conditions and formation processes of the
early Universe, providing important constraints which must be
reproduced in cosmological models. It is therefore valuable to fully
outline these superclusters across the ZOA.

It is curious, that the two major superclusters in the local Universe,
\ie Perseus-Pisces and the Great Attractor overdensity, lie at similar
distances on opposite sides of the LG, and that both are partially obscured
by the ZOA. It is therefore of particular interest to map these
structures in detail, determine their extent and masses, in order to
find out which one of the two is dominant in the tug-of-war on the
Local Group.

\subsection{Unveiling Large-Scale Structures Behind the Milky Way}

For all of the above reasons, the unveiling of galaxies behind the
Milky Way has turned into a research field of its own in the last ten
years. In the following, I discuss all the various observational
multi-wavelength techniques that are currently being employed to
uncover the galaxy distribution in the ZOA such as deep optical
searches, far-infrared and near-infrared surveys, systematic blind
radio surveys and searches for hidden massive X-ray clusters. I will
describe the different limitations and selection effects inherent to
each method and present results obtained with these various methods --
describing the results and discoveries in detail for the Great
Attractor region. Predictions from reconstructions of the density
field in the ZOA are also presented and compared with observational
evidence. The comparison between reconstructed density fields and the
observed galaxy distribution are important as they allow derivations
of the density and biasing parameters $\Omega_0$ and $b$.

\section{Optical Galaxy Searches} \label{opt}

Systematic optical galaxy catalogs are generally limited to the
largest galaxies (typically with diameters $D \GA 1\arcmin$, \eg\
\cite{Lau82}). 
These catalogs become, however, increasingly
incomplete for galaxies the closer they are to the Galactic Plane.
With the thickening of the dust layer, the absorption increases and
reduces the brightness of the galaxies and their `visible' extension.
Obviously such galaxies are not intrinsically faint; they only appear
faint because of the dimming by the dust. Systematical deeper
searches for partially obscured galaxies -- down to fainter magnitudes
and smaller dimensions compared to existing catalogs -- have been
performed on sky surveys with the aim of reducing this ZOA.

\subsection{Early Searches and Results} \label{early}

One of the first attempts to detect galaxies in the ZOA was carried
out by B\"ohm-Vitense in 1956 \cite{Boh56}. 
She did follow-up observations in
selected fields in the GP in which Shane \& Wirtanen \cite{Sha54} 
found
objects that "looked like extragalactic nebulae" but were not believed
to be galaxies because they were so close to the dust equator. She
confirmed many galaxies and concluded that the obscuring matter in the
plane must be extremely thin and full of holes between
$\ell = 125\deg$-$130\deg$.

Because extinction was known to be low in Puppis, Fitzgerald \cite{Fit74} 
performed a galaxy search on a field there ($\ell \sim 245\deg$) and
discovered 18 small and faint galaxies. Two years later, Dodd \& Brand
\cite{Dod76} 
examined 3 fields adjacent to this area ($\ell \sim 243\deg$)
and detected another 29 galaxies.  Kraan-Korteweg \& Huchtmeier 
\cite{Kra92} 
observed these galaxies at radio wavelengths with the 100~m
radio telescope at Effelsberg in Germany. This method was chosen because
extinction is unimportant at these long wavelengths and the neutral
gas of spiral galaxies can easily be observed at 21~cm (see
Sect.~\ref{hi}). With these observations, a previously unknown nearby
cluster at ($\ell,b,v) = (245\deg,0\deg, \sim1500$~\kms) could be
identified. Adding far-infrared data (see Sect.~\ref{fir}), it was
shown that this Puppis cluster is comparable to the Virgo cluster and
that it contributes a significant component to the peculiar motion of
the LG \cite{Lah93}. 

During a search for infrared objects Weinberger \etal \cite{Wei76}, 
detected
two galaxy candidates near the Galactic Plane ($\ell \sim 88\deg$)
which Huchra \etal \cite{Huc77} confirmed in 1977 to be the brightest members of a
galaxy cluster at 4200~\kms. This discovery led Weinberger \cite{Wei80} 
to start the first {\sl systematic} galaxy search. Using the red prints
of the Palomar Sky Survey, he covered the whole northern GP ($\ell =
33\deg$-$213\deg$) in a thin strip $(|b| \le 2\deg)$. He found 207
galaxies, the distribution of which is highly irregular: large areas
disclose no galaxies, the "hole" pointed out by B\"ohm-Vitense was
verified, but most conspicuous was a huge excess of galaxies around
$\ell = 160\deg$-$165\deg$. In 1984, Focardi \etal \cite{Foc84} 
made the connection
with large-scale structures: they interpreted the excess as the
possible continuation of the Perseus-Pisces cluster [PP] across the
plane to the cluster A569.  Radio-redshift measurements by Hauschildt
\cite{Hau87} 
established that the PP cluster at a mean redshift of $v =
5500$~\kms\ extends to the cluster 3C129 in the GP ($\ell = 160\deg, b
= 0\fdg1$). Additional \HI\ and optical redshift measurements of
Zwicky galaxies by Chamaraux \etal \cite{Cham90} 
indicate that this chain can
be followed even further to the A569 cloud at $v \sim 6000$~\kms\ on
the other side of the ZOA.

These early searches proved that large-scale structure can be traced
to very low Galactic latitudes despite the foreground obscuration
and its patchy nature which shows clumpiness and clustering in the
galaxy distribution independent of large-scale structure. The above
investigations did confirm suspected large-scale features across the
plane through searches in selected regions and follow-up redshift
observations. To study large-scale structure, systematically
broader latitude strips covering the whole Milky Way, respectively
the whole ZOA (see Fig.~\ref{ait}) are required.

\subsection{Status of Systematic Optical Searches}\label{optsear}

Using existing sky surveys such as the first and second generation
Palomar Observatory Sky Surveys POSS I and POSS II in the north, and
the ESO/SRC (United Kingdom Science Research Council) Southern Sky
Atlas, various groups have performed systematic deep searches for
``partially obscured'' galaxies. They catalogued galaxies down to
fainter magnitudes and smaller dimensions (${D} \GA 0\farcm1$) than
previous catalogs. Here, examination by eye remains the best
technique. A separation of galaxy and star images can as yet not be
done on a viable basis below $|b| \LA 10\deg$-$15\deg$ by automated
measuring machines such as \eg COSMOS \cite{Dri95} 
or APM \cite{Lew96} 
and sophisticated extraction algorithms, nor
with the application of Artificial Neural Networks. Thus,
although surveys by eye clearly are both very trying and time
consuming -- and maybe not as objective -- they currently still
provide the best technique to identify partially obscured galaxies in
crowded star fields.

Meanwhile, through the efforts of various collaborations, nearly the
whole ZOA has been surveyed and over 50000 previously unknown galaxies
could be discovered in this way. These surveys are not biased with
respect to any particular morphological type. The various surveyed
regions are displayed in Fig.~\ref{cor}. Details and results on the
uncovered galaxy distributions can be found in the respective
references listed below:

\begin{figure}[ht]
\begin{center}
\includegraphics[width=12cm]{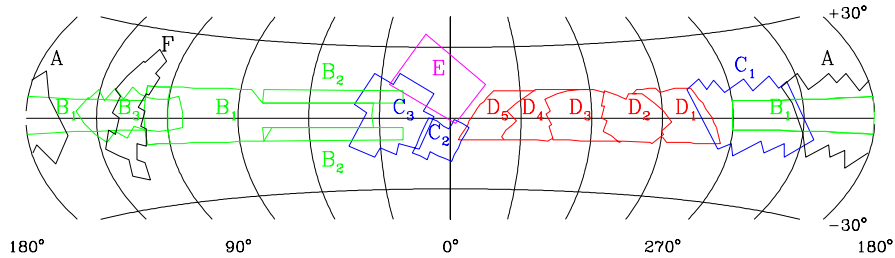}
\caption
{An overview of the different optical galaxy surveys in the ZOA centered
on the Galaxy. The labels identifying the search areas are explained 
in the text. Note that the surveyed regions cover the entire ZOA as
defined by the foreground extinction level of ${A_B} = 1\fm0$
displayed in Fig.~\ref{ait}}
\label{cor}
\end{center}
\end{figure}

\noindent
{\bf A}:  the Perseus-Pisces Supercluster by Pantoja \cite{Pan97}; 
{\bf B$_{1-3}$}: the northern Milky Way (B$_1$ by Seeberger \etal 
\cite{See94,See96,See98}, 
Lercher \etal \cite{Ler96}, 
and Saurer \etal \cite{Sau97}, 
from POSS I; B$_2$ by Marchiotto \etal 
\cite{Mar99} 
also from POSS II; B$_3$ by Weinberger \etal \cite{Wei99} 
from POSS II);\\
{\bf C$_{1-3}$}: the Puppis region  by Saito \etal \cite{Sai90,Sai91}
[C$_1$], the Sagittarius/Galactic region by Roman \etal \cite{Rom98} 
[C$_2$], and the Aquila and Sagittarius region by Roman \etal \cite{Rom96} 
[C$_{3}$]; \\ 
{\bf D$_{1-5}$}: the southern Milky Way (the Hydra to Puppis region [D$_1$]
by Salem \& Kraan-Korteweg \cite{Sal00}, 
the Hydra/Antlia Supercluster
region [D$_2$] by Kraan-Korteweg \cite{Kra00}, 
the Crux region [D$_3$] by 
Woudt \cite{Wou98}, 
Woudt \& Kraan-Korteweg \cite{Wou00a}, 
the GA region [D$_4$] by Woudt \cite{Wou98}, 
Woudt \& Kraan-Korteweg \cite{Wou00b}, 
and the Scorpius region 
[D$_5$] by Fairall \& Kraan-Korteweg \cite{Fai00}; 
{\bf E}:  the Ophiuchus Supercluster by Wakamatsu \etal \cite{Wak94},
Hasegawa \etal \cite{Has00};
{\bf F}: the northern GP/SGP crossing by Hau \etal \cite{Hau95}. 

Comparing the surveyed regions (Fig.~\ref{cor}) with the ZOA as
outlined in Fig.~\ref{ait} clearly demonstrates that nearly the whole
ZOA has been covered by systematic deep optical galaxy searches. 

\subsection{The Galaxy Distribution in the Great Attractor Region}

Most of these searches have quite similar characteristics. As an
example, I discuss in the following the optical galaxy search
performed by our group in the Great Attractor region (D$_{1-5}$).

The tools for this galaxy search were simple. It comprised a viewer
with the ability to magnify 50 times and the IIIaJ film copies of the
ESO/SRC survey.  The viewer projects an area of $3\farcm5 \times
4\farcm0$ on a screen, making the visual, systematic scanning of these
plates quite straightforward and comfortable.
 
Even though Galactic extinction effects are stronger in the blue, the
IIIaJ films were searched rather than their red counterparts. Comparison
between the various surveys demonstrated that the hypersensitized and
fine grained emulsion of the IIIaJ films go deeper and show higher
resolution.  Even in the deepest extinction layers of the ZOA, the red
films were found to have no advantage over the IIIaJ films.

\begin{figure}[ht]
\begin{center}
\includegraphics[width=12cm]{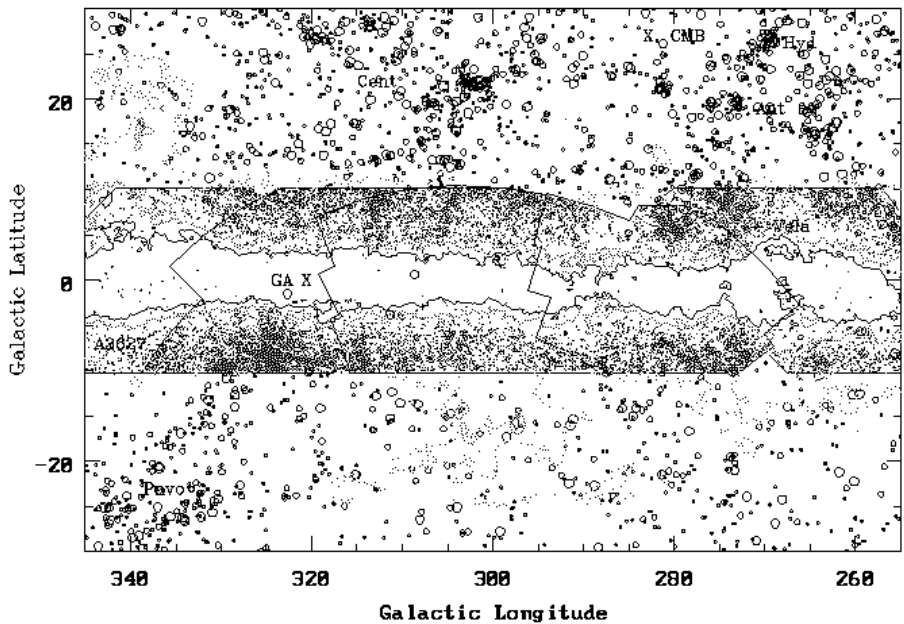} 
\caption
{Distribution of Lauberts galaxies with $D \ge 1\farcm3$ (open circles
-- coded as in Fig.~\ref{ait}) and galaxies with $D \ge 12\arcsec$
(small dots) identified in the deep optical galaxy searches
D$_1$-D$_5$. The contours represent extinction levels of $A_B =
1\fm0$, $3\fm0$ and $5\fm0$. Note how the ZOA could be filled to $A_B =
3\fm0$ and that galaxy over- and underdensities uncorrelated with 
extinction can be recognized in this distribution}
\label{WKK}
\end{center}
\end{figure}

A diameter limit of $D \GA 0\farcm2$ was imposed. Below this diameter
the reflection crosses of the stars disappear, making it hard to
differentiate consistently between stars or blended stars and faint
galaxies. The positions of all the galaxies are measured with the
Optronics, a high precision measuring machine, at ESO (European
Southern Observatories) in Garching, Germany. The accuracy of these
positions is about $1\arcsec$.  For every galaxy we recorded the major
and minor diameter, an estimate of the average surface brightness and
the morphological type of the galaxy. From the diameters and the
average surface brightness a magnitude estimate was derived. A
surprisingly good relation was found for the estimated magnitudes, with
no deviations from linearity even for the faintest galaxies, and a
scatter of only $\sigma = 0\fm5$ \cite{Kra00}. 
In this
manner over 17\,000 galaxies in about 1800~sq. deg. could be
identified, of which $\sim$ 97\% were previously unknown. Their
distribution is displayed in Fig.~\ref{WKK} together with all the
Lauberts galaxies larger than $D \ge 1\farcm3$ (diameter-coded as in
Fig.~\ref{ait}) as well as the DIRBE foreground extinction contours of
$A_B = 1\fm0$, $3\fm0$ and $5\fm0$.

The distribution reveals that galaxies can easily be traced through
obscuration layers of 3 magnitudes, thereby narrowing the ZOA
considerably. A few galaxies are still recognizable up to extinction
levels of $A_B = 5\fm0$ and a handful of very small galaxy
candidates have been found at even higher extinction levels. The
latter most likely indicate holes in the dust layer.  Overall, the
mean number density follows the dust distribution remarkably well at
low Galactic latitudes. The contour level of $A_B = 5\fm0$, for
instance, is nearly indistinguishable from the galaxy density contour
at 0.5 galaxies per square degree.

At intermediate extinction levels (between the outer and second
extinction contour $1\fm0 \le A_B \le 3\fm0$), distinct under- and
overdensities are noticeable in the unveiled galaxy distribution that
are uncorrelated with the foreground obscuration.  They must be the
signature of large-scale structures.

The most extreme overdensity is found at $(\ell,b) \sim
(325\deg,-7\deg$). It is at least a factor 10 denser compared to
regions at similar extinction levels. This galaxy excess is centered
on the cluster A3627. It is the only cluster out of 4076 clusters in
the Abell cluster catalog \cite{Abe89}.
Although it is (a) classified as a rich, nearby cluster, (b) the only
Abell cluster identified below $|b| < 10\deg$, and (c) within a few
degrees of the predicted center of the GA \cite{Kol95}, 
this cluster had not received any attention.  This is mainly due to the
foreground obscuration. A3627 is hardly discernable in, for instance,
the distribution of Lauberts galaxies: the observed diameters of the
galaxies in this density peak are just {\sl below} the Lauberts
diameter limit (due to the obscuration). This cluster is {\sl not
evident} in the far infrared (see Sect.~\ref{fir}). This can be
explained by the predominance of early-type galaxies (50\% in the core
of this cluster, 25\% within its Abell radius) which do not radiate in
the far infrared but are a clear signature of rich clusters. The new
data support the classification of A3627 as a rich cluster: over 600
likely new cluster members were identified compared to the 50 larger
galaxies noted by Abell.

\begin{figure}[ht]
\begin{center}
\includegraphics[width=10cm]{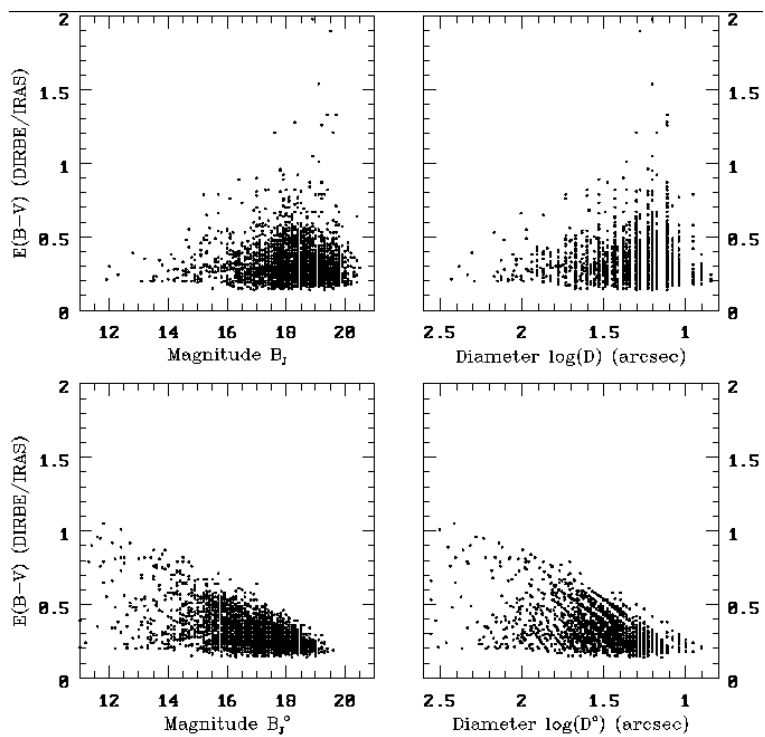}
\caption
{The observed (top panels) and extinction-corrected (bottom)
magnitudes (left) and diameters (right) of galaxy candidates in the
Hydra/Antlia region as a function of the foreground extinction $E(B-V)$}
\label{ebv}
\end{center}
\end{figure}

The galaxies detected in these searches are quite small ($<D> =
0\farcm4$) and faint ($<B_J> = 18\fm0$) on average. So the question
arises whether these new galaxies and the newly uncovered over- and
underdensities are relevant at all to our understanding of the
dynamics in the local Unverse. To assess this, we have to understand
the effects of extinction: galaxies are diminished by at least $1^{\rm
m}$ of foreground extinction at the highest latitudes ($|b| \sim
10\deg$) of the search areas. These effects increase considerably
closer to the Galactic Equator. The effects of the absorption on the
observed parameters of these low-latitude galaxies is reflected
clearly in Fig.~\ref{ebv}. Here, the magnitudes and major diameters of
galaxies in the Hydra/Antlia search region (D$_2$) are plotted against
the Galactic extinction $E(B-V)$ derived from the 100 micron DIRBE
dust maps \cite{Sch98}. 
The top panels show the observed
magnitudes (left) and diameters (right).

The distribution of both the observed magnitudes and diameters show a
distinct cut-off as a function of extinction -- all the galaxies lie
in the lower right triangle of the diagram, leaving the upper left
triangle empty. At low extinction values, bright and faint galaxies can
be identified, whereas apparently faint and small galaxies remain
visible only at higher extinction values. The division in the diagram
defines an upper envelope of the intrinsically brightest and largest
galaxies. This fiducial line, i.e. the shift $\Delta m$ to fainter
apparent magnitudes of the intrinisically brightest galaxies, is a
direct measure of the absorption ${A}_{B}$. In fact, this shift in
magnitude is tightly correlated with the absorption in the blue
${A}_{B} = 4.14 \cdot {E(B-V)}$. The galaxies at these extinction
levels are not intrinsically faint. They must in fact be intrinsically
very bright galaxies to still be visible through the murk of the Milky Way.

The obscuration effects on the parameters of galaxies have been
studied in detail by Cameron \cite{Cam90} 
who simulated the effects of
absorption on the brightness profiles of various Virgo
galaxies. This led to analytical descriptions of the diameter and
isophotal magnitude corrections given in Table~\ref{cameron} for
early-type and spiral galaxies:

\begin{center}
\begin{table}[h]
 \caption{Obscurational effects on the diameter and isophotal magnitude.}
 \begin{center}
 \begin{tabular}{lcc}
   & Reduction factor & Additional $\Delta$m \\  
\hline 
 ellipticals/lenticulars & $10^{0.13 {A_B}^{1.3}}$ & $0.08 {A_B}^{1.8}$ \\
\vspace{0.1mm}
 spirals                & $10^{0.10 {A_B}^{1.7}}$ & $0.07 {A_B}^{2.5}$ \\
 \hline
 \end{tabular}
\end{center}
 \label{cameron}
\end{table}
\end{center}

For example, a spiral galaxy, seen through an extinction of $A_B =
1^{\rm m}$, is reduced to $\sim 80\%$ of its unobscured size.  Only
$\sim 22\%$ of a (spiral) galaxy's original dimension is seen when it
is observed through $A_B = 3^{\rm m}$, and its isophotal magnitude
will be diminished by $4\fm1$. Applying these corrections to the
optical ZOA galaxy samples invert the trends in the magnitude and
diameter distributions. This can be verified in the lower panels of
Fig.~\ref{ebv} where the extinction-corrected magnitudes and diameters
are plotted.  At high extinction only the intrinsically bright galaxies
can be identified. These deep optical galaxy searches hence do uncover
intrinsically bright galaxies at lower latitudes.

Correcting the galaxies identified in deep optical searches for
absorption partially lifts the veil of the Milky Way. Without the
extinction layer, the Lauberts catalog would have, for instance, found
139 galaxies with $D \ge 1\farcm0$ within the Abell radius $R_A =
3\,h_{50}^{-1}$\,Mpc for A3627 compared to the previously identified
31 galaxies, where $h_{50}$, the dimensionless Hubble parameter is 1
for a Hubble constant of $H_0 = 50$\,\kms\,Mpc$^{-1}$ ($H_0 =
50\,h$\,\kms\,Mpc$^{-1}$).  This makes this cluster {\sl the most
prominent overdensity in the southern sky}. Were it not for the
obscuration, it most likely would have been the best-studied cluster
in the Universe.

\subsection{Redshift Follow-ups and the Cluster A3627}\label{optred}

Analazing the galaxy density as a function of the galaxy size,
magnitude and/or morphology in combination with the foreground
extinction has led to the identification of various important
large-scale structures in the ZOA and their approximate distances.
Redshift observations must be obtained to map the large-scale
structures in redshift space. So far, this has been pursued
extensively in the Perseus-Pisces supercluster \cite{Pan97},
the Puppis region \cite{Cham99}, 
the Ophiuchus supercluster
behind the Galactic Bulge area \cite{Has00} 
and the southern
ZOA.  Here again, I concentrate on the results from various observing
programs in the Great Attractor region. For a listing of the mapping
of other large-scale structures and references see Kraan-Korteweg \&
Woudt \cite{Kra99}. 
 
For the survey regions D$_{1-5}$ we use complementary observing
approaches to obtain the redshifts (see \cite{Kra94a} 
for a more detailed description):

-- multifiber spectroscopy with the MEFOS instrument \cite{Fel97}
at the 3.6m telescope of ESO. This instrument has the ability to
obtain 29 spectra simultaneously within a one-degree circular field;
ideally suited to probe the densest regions in the uncovered
galaxy distribution,

-- individual spectroscopy of all the brighter galaxies ($B_J \sim
17\fm0 - 17\fm5$, depending on the central surface brightness of a
galaxy) with the 1.9m telescope of the South African Astronomical
Observatory (SAAO) \cite{Kra95,Fai98b,Wou99}.
This method allows homogeneous coverage over the whole search area,
-- 21cm observations of extended, low surface-brightness spiral
galaxies with the 64m radio telescope in Parkes, Australia \cite{Kra97}.
The radio observations are an important
addition as it is impossible to obtain good signal-to-noise optical
spectra for highly obscured low-surface brightness galaxies whereas
the 21cm radiation is not influenced by the dust.

\begin{figure}[ht]
\begin{center}
\includegraphics[width=5.9cm]{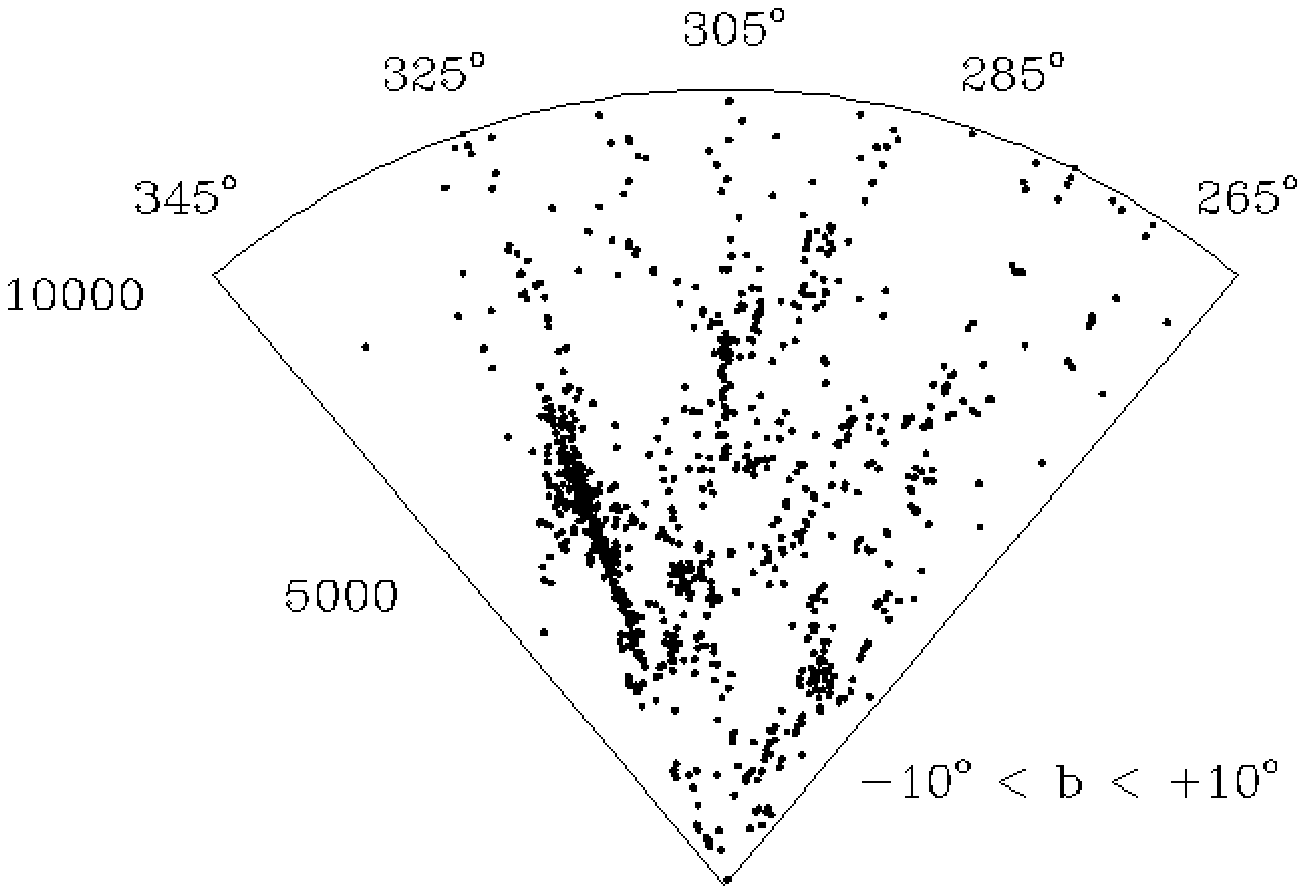} \hspace{-0.5cm}
\includegraphics[width=6.6cm]{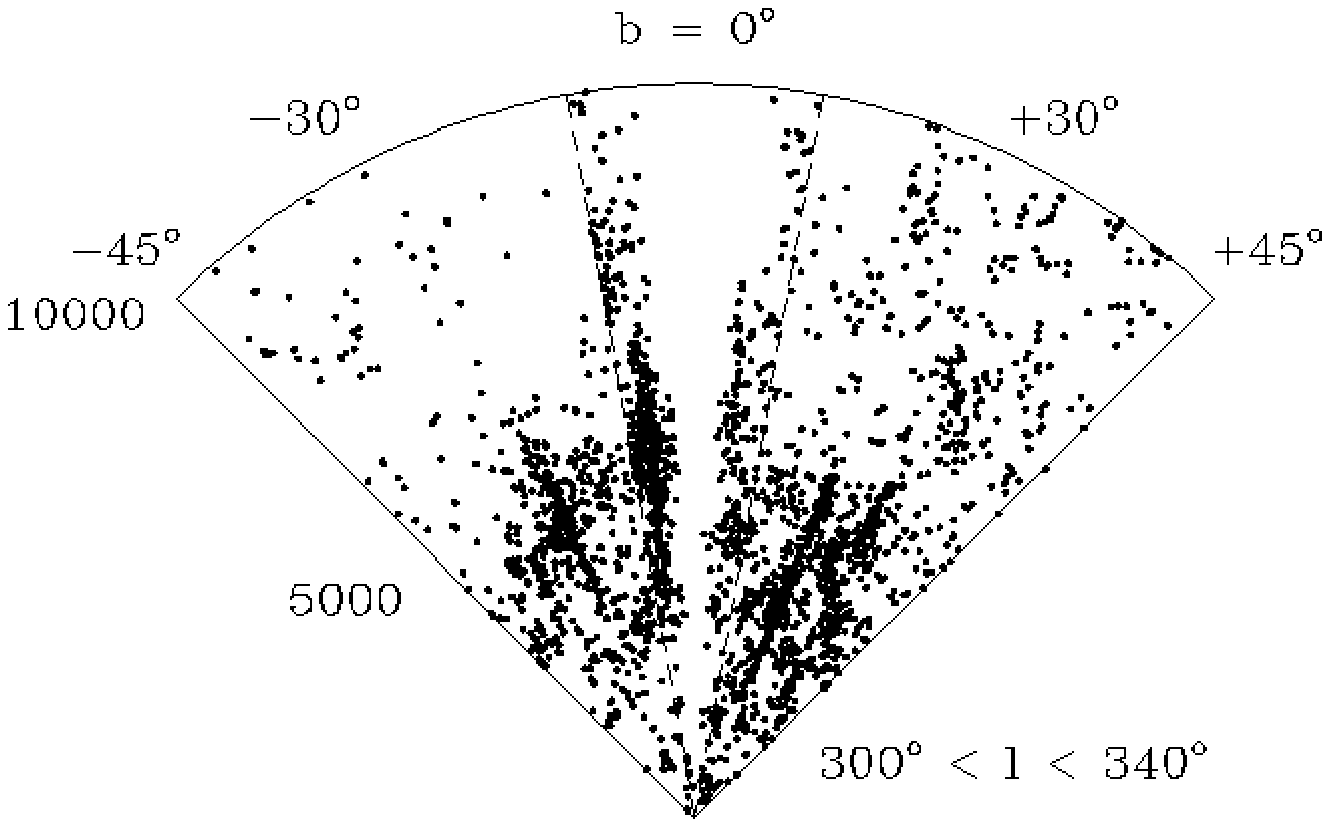} 
\caption
{Redshift slices out to 10000~\kms. The left panel shows the
distribution ``in'' the ZOA ($|b| \protect\LA 10\deg$) along 
Galactic longitudes, the right panel the distribution in the GA region 
($300\deg < \ell < 340\deg$) 
for the latitude range $|b| \le 45\deg$}
\label{pie}
\end{center}
\end{figure}

With the above observations, we typically obtain redshifts of $\GA
10\%$ of the galaxies and can trace large-scale structures out to
recession velocities of $\sim~25000$ \kms. To focus again on the GA
region, a redshift ``slice'' (the distribution of a certain region on
the sky as a function of redshift) out to 10000~\kms\ is shown in the 
left-hand panel of Fig.~\ref{pie} for our optical survey region 
($260\deg \LA \ell \LA 350\deg$, $|b| \LA 10\deg$): a region that
previously was largely blank now reveals clusters, superclusters and
voids. In this illustration, the ZOA is now comparable to other
unobscured regions of the sky. The radially very extended feature at
$\ell=325\deg$ -- the location of the cluster A3627 -- is the
signature of a galaxy cluster: the ``finger of God'' feature due to the
velocity dispersion of a virially bound cluster.

\begin{figure}[ht]
\begin{center}
\includegraphics[width=10cm]{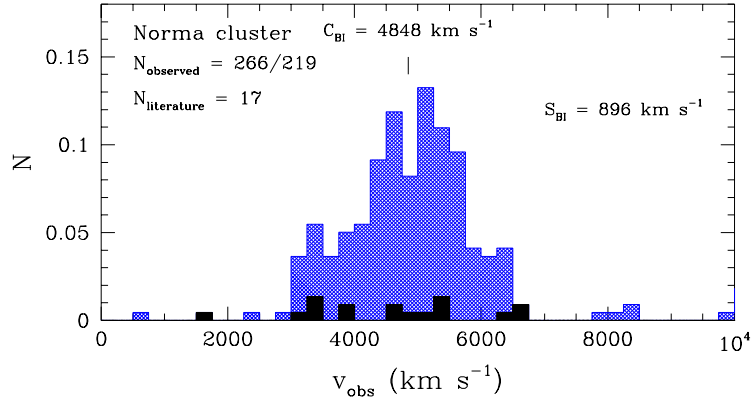}
\caption
{The velocity histogram of galaxies within the Abell radius (${R_A} =
3\,{h}_{50}^{-1}$\,Mpc) of the Norma cluster. Galaxies with redshift
information available in the literature before the ZOA redshift survey
are indicated by the dark shaded histogram. A total of 219 likely
cluster members are identified}
\label{veldis}
\end{center}
\end{figure}

On the right-hand panel, all structures within the general GA region
($300\deg \le \ell \le 340\deg$) are displayed with structures
adjacent to the Milky Way ($-45\deg \le b \le 45\deg$).  Here we can
clearly discern the Hydra ($b=27\deg$), Antlia ($b=19\deg$) and
bimodal Centaurus clusters on the northern side of the Galactic Plane
and the Pavo cluster ($-24\deg$) on the southern side. It is
impressive to note that the new redshifts in the A3627 cluster area
prove this cluster to be the dominant structure within the general GA
overdensity.  While this cluster includes the well-researched radio
galaxy PKS1610$-$601, relatively few redshifts of other cluster
members were known beforehand. Adding, however, the new ZOA redshift
data, we find a near Gaussian distribution of the velocities, resulting
in a mean observed velocity of $<v>=4848$~\kms\ and a velocity
dispersion of $\sigma = 896$~\kms. This is displayed in
Fig.~\ref{veldis} where the dark shaded histogram identifies
previously known galaxies and the light shaded histogram the redshift
data from our ZOA program.

The large dispersion suggests A3627 to be a massive cluster. The
dynamical mass within a radius $R$ \cite{Sar86} 
is given by
$${\cal M} (< R) = {{9 \sigma^{2} R_c} \over {{\rm G}}} (\ln (x +
(1+x^2)^{1/2}) - x ( 1 + x^2)^{-1/2})$$
where $\sigma$ is the measured line-of-sight velocity dispersion
(corrected for the errors in the velocity measurements), $R_c$ is the
core radius \cite{Kin62}, 
G is the gravitational constant, and $x = R/R_c$.

With a core radius of 0.29 ${h_{50}^{-1}}$~Mpc, a virial mass within
the Abell radius $R_A = 3 {h}_{50}^{-1}$~Mpc of
$${\cal M}_{A3627} = 0.9 \cdot 10^{14} h_{50}^{-1}{\cal M_\odot}$$ is
found for A3627. This mass is typical of rich clusters, and
comparable, for instance, to the well-studied Coma cluster 
\cite{Hug90,Whi93}.
The latter was already identified in 1906 by
Wolf \cite{Wol06} in the distribution of nebulae (galactic and extragalactic).
With a mean redshift of 6960~\kms, the Coma cluster counted as the
nearest rich cluster. At a mean redshift of 4848~\kms, this place is
now being usurped by the A3627 cluster, also called Norma cluster for
the constellation it lies in.

Rich massive clusters generally are strong X-ray emitters (see
Sect.~\ref{Xray}) and were identified early on with X-ray satellites
(Einstein, HEAO, Uhuru) -- except for A3627. However, A3627 was
detected in a whole-sky survey by the X-ray satellite ROSAT, in which
the Norma cluster ranks as the 6$^{th}$ brightest X-ray cluster in the
sky compared to Coma, which ranks 4 \cite{Boh96}. 

The mean velocity of the Norma cluster puts it well within the
predicted velocity range of the GA. Including the new results
from the deep optical galaxy search, the Norma cluster now is the most
massive galaxy cluster in the GA region known to date.  It most likely
marks the previously unidentified but predicted density-peak at the
bottom of the potential well of the GA overdensity.

The mass excess of the GA is presumed to arise within an area of
radius of about 20~Mpc \cite{Lyn91}. 
These extended
potential wells generally have a rich cluster at their center.  This
actually matches the emerging picture quite well: A3627 appears to lie
at the center of an apparent ``great wall''-like structure, similar to Coma
in the (northern) Great Wall. The right-hand redshift slice of
Fig.~\ref{pie} suggests a very large-scale coherent structure, starting
at Pavo ($332\deg,-24\deg$) and moving towards the density peak of A3627
at slightly larger velocities. This supercluster then seems to
bend towards or merge with the Vela supercluster at $(l,b,v) \sim
(280\deg,6\deg,\sim6000$~\kms) postulated by Kraan-Korteweg 
\etal \cite{Kra94a}. 

One can, however, not exclude the possibility that other unknown rich
clusters reside in the GA region, as the ZOA has not been fully mapped
with the optical galaxy searches (see Fig.~\ref{WKK} and right panel
of Fig.~\ref{pie}). Finding a further uncharted, rich cluster of
galaxies at the heart of the GA would have serious implications for
our current understanding of this massive overdensity in the local
Universe.  Various indications suggest, for instance, that
PKS1343$-$601, the second brightest extragalactic radio source in the
southern sky, might form the center of yet another highly obscured
rich cluster \cite{Kra99}, 
particularly as it also
shows significant X-ray emission. At ($\ell, b) \sim (310\deg,2\deg)$,
this radio galaxy lies behind an obscuration layer of about 12
magnitudes of extinction in the B-band, hence optical surveys are
ineffective.  Still, West \& Tarenghi observed this source in 1989 
\cite{Wes89}:
with an extinction-corrected diameter of ${D^o} \sim 4\arcmin$ and a
recession velocity of $v = 3872$~\kms\ this galaxy appears to be a
giant elliptical galaxy and giant ellipticals are mainl found at the
cores of clusters.

\begin{figure}[ht]
\begin{center}
\includegraphics[width=12cm]{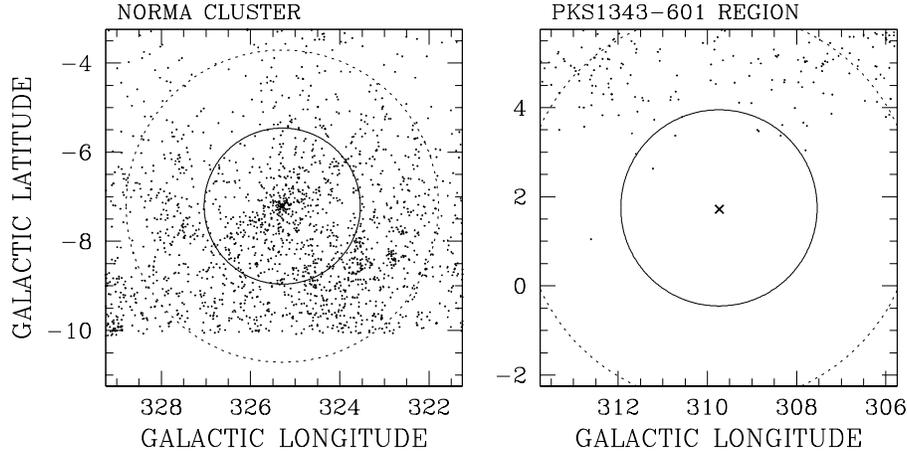}
\caption
{Sky distribution of galaxies identified in the deep optical galaxy 
search around the rich A3627 cluster ($A_{B}\sim1\fm5$) and around
the suspected cluster centered on PKS1343$-$601 ($A_{B}\sim 12^{m}$), both in
the GA region. The inner circle marks the Abell radius $R_{A} = 
3\,h_{50}^{-1}$~Mpc}
\label{pks1343}
\end{center}
\end{figure}

Since PKS1343$-$601 is so heavily obscured, little data are
available to substantiate the existence of this prospective cluster.  In
Fig.~\ref{pks1343} the A3627 cluster at a mean extinction $A_B =
1\fm5$ as seen in deep optical searches is compared to the prospective
PKS1343 cluster at ($309\fdg7, +1\fdg7, 3872$~\kms) with an extinction
of 12$^{\rm m}$.  One can clearly see, that at the low Galactic
latitude of the suspected cluster PKS1343, the optical galaxy survey
could not retrieve the underlying galaxy distribution, especially not
within the Abell radius of the suspected cluster (the inner circle in
the right panel of Fig.~\ref{pks1343}). To verify this cluster, other
observational approaches are necessary. Interestingly enough, deep
\HI\ observations did uncover a significant excess of galaxies at this
position in velocity space (see Sect.~\ref{MBdeep}) although a
``finger of God'', the characteristic signature of a cluster in
redshift space, is not seen.  Hence, the Norma cluster A3627 remains
the best candidate for the center of the extended GA overdensity.

\subsection{Completeness of Optical Galaxy Searches}\label{optcomp}

In order to merge the various deep optical ZOA surveys with existing
galaxy catalogs, Kraan-Korteweg \cite{Kra00} 
and Woudt \cite{Wou98} 
have analyzed
the completeness of their ZOA galaxy catalogs as a function of the
foreground extinction. By studying the apparent diameter distribution
as a function of the extinction, as shown in Fig.~\ref{ebv}, as well
as the location of the flattening in the slope of the cumulative
observed and extinction-corrected diameter curves $(\log D) - (\log
N)$ and $(\log D^o) - (\log N)$ for various extinction intervals (\cf
Fig.~6 in \cite{Kra00}), 
they concluded that the optical ZOA
surveys are complete to an apparent diameter of ${D} = 14\arcsec$ --
where the diameters correspond to an isphote of 24.5~mag/arcsec${^2}$
-- for extinction levels less than $A_{B} = 3\fm0$ (see also
Fig.~\ref{ebv}).

What about the intrinsic diameters, \ie the diameters galaxies would
have if they were unobscured?  Applying the Cameron corrections, it
was found that at $A_{B} = 3\fm0$, an obscured spiral or an elliptical
galaxy at the completeness limit ${D} = 14\arcsec$ would have an
intrinsic diameter of ${D^o} \sim 60\arcsec$, respectively ${D^o} \sim
50\arcsec$. At extinction levels higher than $A_{B} = 3\fm0$, an
elliptical galaxy with $D^o = 60\arcsec$ would appear smaller than the
completeness limit $D = 14\arcsec$ and might have gone unnoticed.
These optical galaxy catalogs should therefore be complete to $D^o \ge
60\arcsec$ for all galaxy types down to extinction levels of $A_{B}
\le 3\fm0$, with the possible exception of extremely low-surface
brightness galaxies. Only intrinsically very large and bright galaxies
-- particularly galaxies with high surface brightness -- will be
recovered in deeper extinction layers.  This completeness limit could
be confirmed by independently analyzing the diameter vs. extinction
and the cumulative diameter diagrams for extinction-corrected
diameters.

We can thus supplement the ESO, UGC and MCG catalogs (see
Fig.~\ref{ait}), which are complete to ${D} = 1\farcm3$, with
galaxies from optical ZOA galaxy searches that have ${D^o} \ge
1\farcm3$ and ${A_B} \le 3\fm0$. As our completeness limit lies
well above the ESO, UGC and MCG catalogs, we can assume that the other
similarly performed optical galaxy searches in the ZOA should also be
complete to ${D^o} = 1\farcm3$ for extinction levels of ${A_B}
\le 3\fm0$.

\begin{figure}
\begin{center}
\includegraphics[width=12cm]{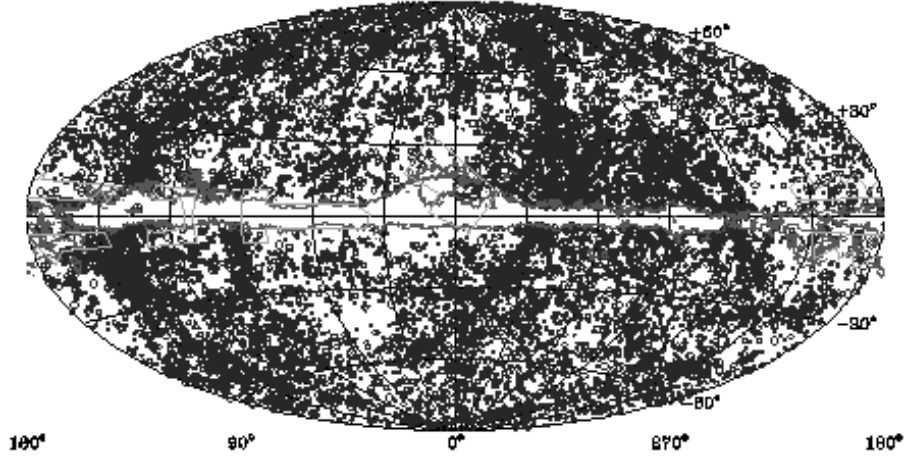}
\caption
{Aitoff equal-area distribution in Galactic coordinates 
of ESO, UGC, MCG galaxies with 
extinction-corrected diameters ${D^o} \ge 1\farcm3$, including 
galaxies identified in the optical ZOA galaxy searches for 
extinction-levels of ${A_B} \le 3\fm0$ (contour). The diameters are
coded as in Fig.~\ref{ait}. With the exception of the areas for
which either the positions of the galaxies or their diameters are
not yet available (demarcated areas), the ZOA could be reduced
considerably compared to Fig.~\ref{ait}}
\label{aitc}
\end{center}
\end{figure}

With Fig.~\ref{aitc}, the first attempt has been made to arrive at an
improved whole-sky galaxy distribution with a reduced ZOA. In this
Aitoff projection all the UGC, ESO, MCG galaxies that have {\it
extinction-corrected} diameters ${D^o} \ge 1\farcm3$ are plotted
[remember that galaxies adjacent to the optical galaxy search regions
are also affected by absorption though to a lesser extent ($A_B \le
1\fm0$)], including the galaxies other optical surveys for which
positions and diameters were available. The regions for which these
data are not yet available are marked in Fig.~\ref{aitc}. As some
searches were performed on older generation POSS I plates, which are
less deep compared to the second generation POSS II and ESO/SRC
plates, an additional correction was applied to those diameters, \ie\
the same correction as for the UGC galaxies which also are based on
POSS I survey material (${D_{25} = 1.15 \cdot D_{\rm POSS\,I}}$).

A comparison of Fig.~\ref{ait} with Fig.~\ref{aitc} demonstrates
convincingly how the deep optical galaxy searches realize a
considerable reduction of the ZOA; we can now trace the large-scale
structures in the nearby Universe to extinction levels of ${A_B} =
3\fm0$. Inspection of Fig.~\ref{aitc} reveals that the galaxy density
enhancement in the GA region is even more pronounced and a connection
of the Perseus-Pisces chain across the Milky Way at $\ell=165\deg$
more likely. Hence, these supplemented whole-sky maps certainly should
improve our understanding of the velocity flow fields and the total
gravitational attraction on the Local Group.

Optical galaxy searches, however, fail in the most opaque part of the
Milky Way, the region encompassed by the ${A_B} = 3\fm0$ contour in
Fig.~\ref{aitc} -- a sufficiently large region to hide further
dynamically important galaxy densities. Here, other systematic surveys
in other wavebands can be applied to reduce the current ZOA even
further. The success and status of these approaches are discussed in
the following sections.
 
\section{Far Infrared Surveys and the ZOA}\label{fir}

In 1983, the Infrared Astronomical Satellite IRAS surveyed 96\% of the
whole sky in the far infrared bands at 12, 25, 60 and 100~$\micron$,
resulting in a catalog of 250\,000 point sources, i.e. the IRAS Point
Source Catalogue \cite{Joi88}.
The latter has been used extensively to quantify extragalactic
large-scale structures.  The identification of the galaxies from the
IRAS data base is quite different compared to the optical: only the
fluxes at the 4 far infrared (FIR) IRAS passbands are available but no
images. The identification of galaxies is strictly based on the
relation of the fluxes. For instance, Yamada \etal \cite{Yam93} 
used the
criteria: {\bf 1.}  $f_{60} > 0.6$Jy, {\bf 2.} $f^2_{60} >
f_{12} f_{25}$, {\bf 3.} $0.8 < f_{100}/f_{60} < 5.0$, to select
galaxy candidates from the IRAS PSC.

With these flux and color criteria mainly normal spiral galaxies
and starburst galaxies are identified. Hardly any dwarf galaxies enter
the IRAS galaxy sample, nor the dustless elliptical galaxies, as
they do not radiate in the far infrared. The upper cut-off in the third
criterion is imposed to minimize the contamination with cool cirrus
sources and young stellar object within our Galaxy. This, however, 
also makes the IRAS surveys less complete for nearby galaxies 
\cite{Wou98,Kra00}.

The advantage of using IRAS data for large-scale structure studies is
its homogeneous sky coverage (all data from one instrument) and the
negligible effect of the extinction on the flux at these long
wavelengths.  Even so, it remains difficult to probe the inner part of
the ZOA with IRAS data because of cirrus, high source counts of
Galactic objects in the Galaxy, and confusion with these objects --
most of them have the same IRAS characteristics as external
galaxies. The difficulty in obtaining unambiguous galaxy
identifications at these latitudes was demonstrated by Lu \etal \cite{Lu90},
who found that the detection rate of IRAS galaxy candidates decreases
strongly as a function of Galactic latitude (from $|b| = 16\deg$ to
$|b| = 2\deg$). This can only be explained by the increase in faulty
IRAS galaxy identifications. Yamada \etal \cite{Yam93} 
also found a dramatic
and unrealistic increase in possible galaxies close to the Galactic
Plane in their systematic IRAS galaxy survey of the southern Milky Way
($|b| \le 15\deg$).

\begin{figure}[ht]
\begin{center}
\includegraphics[angle=-90,width=12cm]{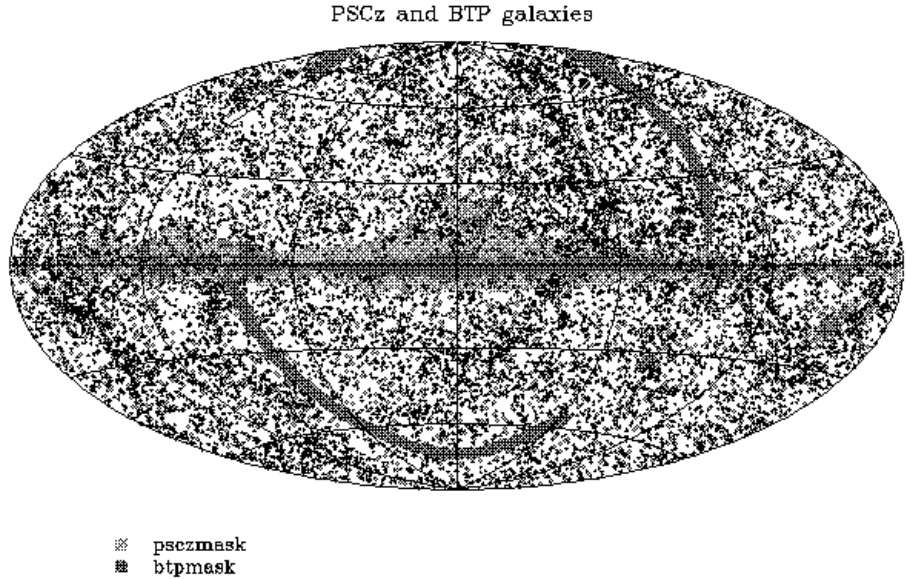}
\caption
{The PSCz and BTP IRAS galaxy catalogs centered on the Galaxy with the
PSCz incompleteness mask (light-grey mask) and the BTP mask
(dark-grey). Note the dramatic reduction of the incompleteness around
the Galactic Equator due to the BTP survey}
\label{BTP}
\end{center}
\end{figure}

So, despite the various advantages given with IRAS data, the sky
coverage in which reliable IRAS galaxy identifications can be made
(84\%) provides only a slight improvement over optical galaxy catalogs
(compare \eg the light-grey mask in Fig.~\ref{BTP} with the optical
ZOA-contour as displayed in Fig.~\ref{ait}). In addition to that, the
density enhancements are very weak in IRAS galaxy samples because (a)
the IRAS luminosity function is very broad, which results in a more
diluted distribution since a larger fraction of distant galaxies will
enter a flux-limited sample compared to an optical galaxy sample, and
(b) IRAS is insensitive to elliptical galaxies, which reside mainly in
galaxy clusters, and mark the peaks in the mass density distribution
of the Universe. This is quite apparent in a comparison of the IRAS
galaxy distribution (Fig.~\ref{BTP}) with the optical galaxy
distribution (Fig.~\ref{ait} and Fig.~\ref{aitc}).

Nevertheless, dedicated searches for large-scale clustering within the
whole ZOA ($|b| \le 15\deg$) have been made by various Japanese
collaborations (see \cite{Tak96} 
for a summary).  They used
IRAS color criteria to select galaxy candidates which were
subsequently verified through visual examination on sky surveys, such
as the POSS of the northern hemisphere and the ESO/SRC for the
southern sky. Because of their verification procedure, this data-set
suffers, however, from the same limitations in highly obscured regions as
optical surveys.

Based on redshift follow-ups of these ZOA IRAS galaxy samples, they
established various filamentary features and connections across the
ZOA. Most coincide with the structures uncovered in optical work. In
the northern Milky Way both crossings of the Perseus-Pisces arms into
the ZOA are very prominent -- considerably stronger in the FIR than at
optical wavelengths -- and they furthermore identified a new structure:
the Cygnus-Lyra filament at ($60\deg-90\deg, 0\deg, 4000$\kms). Across the
southern Milky Way they confirmed the three general concentrations of
galaxies around Puppis ($\ell = 245\deg$), the Hydra-Antlia extension
($\ell = 280\deg$, \cite{Kra95}) 
and the Centauraus
Wall ($\ell = 315\deg$). However, the cluster A3627 is not seen, nor is
the Great Attractor very prominent compared to the optical or to
the POTENT reconstructions described in Sect.~\ref{recon}.

Besides the search for the continuity of structures across the
Galactic Plane, the IRAS galaxy samples have been widely used for the
determination of the peculiar motion of the Local Group, as well as the
reconstructions of large-scale structure across the Galactic Plane
(see Sect.~\ref{recon}). This has been performed on two-dimensional
IRAS galaxy distribution and, in recent years, as well as on their
distribution in redshift space with the availability of redshift
surveys for progressively deeper IRAS galaxy samples, \ie\ 2658
galaxies to f$_{60\mu m} = 1.9$~Jy \cite{Str92}, 
5321 galaxies
to f$_{60\mu m} = 1.2$~Jy \cite{Fish95}, 
and lately the PSCz
catalog of 15411 galaxies complete to f$_{60\mu m} = 0.6$~Jy with 84\%
sky coverage and a depth of 20000~\kms\ \cite{Sau00b}. 

The PSCz is in principal deep enough to see convergence of the
dipole. Saunders and collaborators realized, however, that the 16\% of
the sky missing from the survey causes significant uncertainty,
particularly because of the location behind the Milky Way of many of
the prominent large-scale structures (superclusters as well as
voids). In 1994, they therefore started a longterm program to increase
the sky coverage of the PSCz. Optimizing their color criteria to
minimize contamination by Galactic sources ($f_{60} / f_{25} > 2$,
$f_{60} / f_{12} > 4$, and $1.0 < f_{100}/f_{60} < 5.0$), they
extracted a further 3500 IRAS galaxy candidates at lower Galactic
latitudes (light-grey area of Fig.~\ref{BTP}), reducing the
coverage gap to a mere 7\% (dark-grey area). Taking $K'$ band
snapshots of all the galaxy candidates of their `Behind The Plane'
[BTP] survey, they could add a thousand galaxies to the PSCz sample.

The resulting sky map of 16,400 galaxies (PSCz plus BTP) is shown in
Fig.~\ref{BTP} (from \cite{Sau00a}). 
The BTP survey has reduced
the ``IRAS ZOA'' dramatically.  Some incompleteness remains towards
the Galactic Center, but large-scale structures can easily be identified
across most of the Galactic Plane. In the Great Attractor region, the
galaxies can be traced (for the first time with IRAS data) to the rich
cluster A3627 -- the suspected core of the GA \cite{Kra96}.
The IRAS galaxies overall seem to align well with the Norma
supercluster \cite{Wou00}. 
The BTP collaboration is currently
working hard on obtaining redshifts for these new and heavily obscured
galaxies and exciting new results on large-scale structure across the
Milky Way and dipole determinations can be expected in the near future.

\section{Near Infrared Surveys and the ZOA}\label{nir}

Observations in the near infrared (NIR) can provide important
complementary data to other surveys. With extinction decreasing as a
function of wavelength, NIR photons are up to 10 times less affected
by absorption compared to optical surveys -- an important aspect in the
search and study of galaxies behind the obscuration layer of the Milky
Way. The NIR is sensitive to early-type galaxies -- tracers of
massive groups and clusters -- which are missed in IRAS and \HI\
surveys (Sect.~\ref{fir} and \ref{hi}).  In addition, confusion with
Galactic objects is considerably lower compared to the FIR surveys.
Furthermore, because recent star formation contributes only little
to the NIR flux of galaxies (in contrast to optical and FIR emission),
NIR data give a better estimation of the stellar mass content of
galaxies. 

\subsection{The NIR Surveys DENIS and 2MASS}

Two systematic near infrared surveys are currently being performed.
DENIS, the DEep Near Infrared Southern Sky Survey, is imaging the
southern sky from $-88\deg < \delta < +2\deg$ in the \II\ ($0.8\mu$m),
\J\ ($1.25\mu$m) and \K\ ($2.15\mu$m) bands. 2MASS, the 2 Micron All
Sky Survey, is covering the whole sky in the \J\ ($1.25\mu$m), \HH\
($1.65\mu$m) and \K\ ($2.17\mu$m) bands.  The mapping of the sky is
performed in declination strips, which are $30\deg$ in length and
12~arcmin wide for DENIS, and $6\deg \times 8\farcm5$ for
2MASS. Both the DENIS and 2MASS surveys are expected to complete their
observations by the end of 2000.  The main characteristics of the 2
surveys and their respective completeness limits for extended sources
are given in Table~\ref{nirprop} \cite{Epch97,Epch98,Skr97,Skr98}.

Details and updates on completeness, data releases and
data access for DENIS and 2MASS can be found on the 
websites http://www-denis.iap.fr, and 
http://www.ipac.caltech.edu/2mass, respectively.

\begin{figure} [t]
\includegraphics[width=12cm]{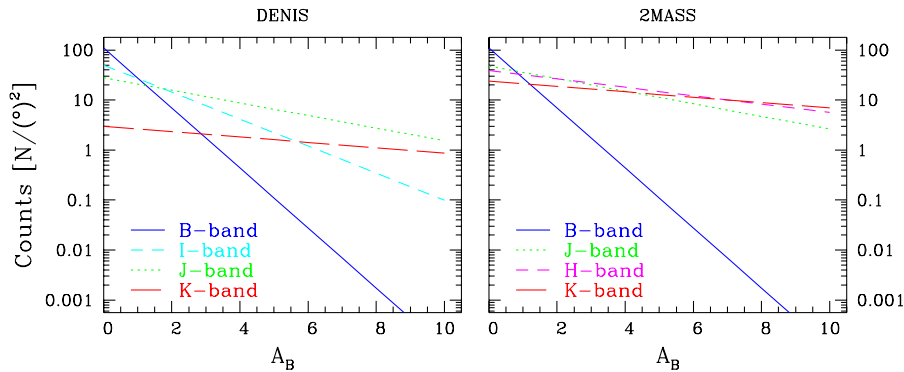}
\caption{Predicted \II, \J\ and \K\ galaxy counts for DENIS (left
 panel), and \J, \HH\ and \K\ counts for 2MASS (right panel) for their
respective galaxy completeness limits as a function of the absorption 
in the \B\ band. For comparison both panels also show the
B counts of an optical galaxy sample extracted from sky surveys
}
\label{nircts}
\end{figure}

The DENIS completeness limits (total magnitudes) for highly reliable
automated galaxy extraction (determined away from the ZOA, \ie $|b| >
10\deg$) are $I = 16\fm5$, $J = 14\fm8$, $K_s = 12\fm0$ \cite{Mam98}.
The number counts per square degrees for these completeness
limits are 50, 28 and 3 respectively.  For 2MASS, the completeness
limits are $J = 15\fm0$, $H = 14\fm2$, $K_s = 13\fm5$ (isophotal
magnitudes), with number counts of 48, $\sim$40 and 24. In all
wavebands, except \II, the number counts are quite imprecise due to
the low number statistics and the strong dependence on the
star crowding in the analyzed fields. Still, they suffice to
reveal the promise of NIR surveys at very low Galactic latitudes.  
As illustrated in Fig.~\ref{nircts}, the galaxy density in 
the \B\ band in unobscured regions is 110 galaxies per square 
degree for the completeness limit of $B_J\le19\fm0$ \cite{Gar96}.
These counts drop rapidly with increasing obscuration: 
$N(A_{B}) \simeq 110 \times {\rm dex} (0.6\,[-A_{B}])\,$deg$^{-2}$. 
The decrease in detectable galaxies due to extinction is much slower 
in the NIR, \ie 45\%, 21\%, 14\% and 9\% compared to the optical for
the \II, \J, \HH\ and \K\ bands. This dependence makes NIR
surveys very powerful at low Galactic latitudes even though they are
not as deep as the POSS and ESO/SRC sky surveys:
the NIR counts of the shallower NIR surveys overtake the optical
counts at extinction levels of $A_B \GA 2$-$3^{\rm m}$.  The location of the
reversal in efficiency is particularly opportune because the NIR
surveys become more efficient where deep optical galaxy searches
become incomplete, \ie at $A_B \GA 3\fm0$ (see Sect.~\ref{optcomp}).

\begin{center}
\begin{table}
\caption{Main characteristics of the DENIS and 2MASS surveys}
\label{nirprop}
\begin{center}
\begin{tabular}{lccccccc}
\vspace{-1mm} \\
& \multicolumn{3}{c}{DENIS}                        &$\;\;\;\;\;\;$&        \multicolumn{3}{c}{2MASS} \\
Channel            &  \II\       & \J\        & \K\        &&   \J\      & \HH\       & \K\        \\
\vspace{-1mm} \\
\hline 
\vspace{-1mm} \\
Central wavelength &  $0.8\mu$m  & $1.25\mu$m & $2.15\mu$m && $1.25\mu$m & $1.65\mu$m & $2.15\mu$m \\   
Arrays             &  1024x1024  &  256x256   &  256x256   &&  256x256   &  256x256   & 256x256    \\
Pixel size         & $1\farcs0$  & $3\farcs0$ & $3\farcs0$ && $2\farcs0$ & $2\farcs0$ & $2\farcs0$   \\
Integration time   &  9s         & 10s        & 10s        && 7.8s       & 7.8s       & 7.8s       \\
Completeness limit &&&&&&\\
for extended sources & $16\fm5$  & $14\fm8$   & $12\fm0$   && $15\fm0$   & $14\fm2$   & $13\fm5$    \\ 
Number counts for the $\;\;$ &&&&&&\\
completeness limits &    50	 &  28        &    3       &&  48     &
$\sim$40       & 24       \\
Extinction compared &&&&&&&\\
to the optical $A_B$ &    0.45	 &  0.21      & 0.09       &&  0.21     &  0.14       & 0.09       \\ 
\vspace{-1mm} \\
\hline 
\end{tabular}
\end{center}
\end{table}
\end{center}

The above predictions do not take into account any dependence on
morphological type, surface brightness, intrinsic color, orientation
and crowding, which may lower the counts of actually detectable
galaxies counts. 

\subsection{Pilot Studies with DENIS Data in the Great Attractor Region}

To compare the above predictions with real data, Schr\"oder \etal 
\cite{Sch97,Sch99} 
and Kraan-Korteweg \etal \cite{Kra98b} 
examined the efficiency of
uncovering galaxies at high extinctions using DENIS images. The
analyzed regions include the rich cluster A3627 ($\ell,b) =
(325\fdg3,-7\fdg2)$ at the heart of the GA (Norma) supercluster  as
well as its suspected extension across the Galactic Plane.


Three high-quality DENIS strips cross the cluster A3627.
The 66 images on these strips that lie within the Abell-radius were 
inspected by eye.  This covers about one-eighth of the
cluster area.  The extinction over the regarded cluster area varies as
$1\fm2 \le A_B \le 2\fm0$.

On these 66 images, 151 galaxies had previously been identified in the
deep optical ZOA galaxy search \cite{Wou00b}. 
Of these, 122 were recovered in the \II, 100 in the \J, and 74 in the \K\
band. Most of the galaxies not re-discovered in \K\ are low surface
brightness spiral galaxies.

Surprisingly, the \J\ band provided better galaxy detection than the
\II\ band.  In the latter, the severe star crowding makes
identification of faint galaxies very difficult. At these extinction
levels, the optical survey does remain the most efficient in {\it
identifying} obscured galaxies.


The search for more obscured galaxies was made in the region $320\deg
\le \ell \le 325\deg$ and $|b|\le 5\deg$, \ie the suspected crossing
of the GA.  Of the 1800 images in that area, 385 of the then available
DENIS images were inspected by eye (308 in \K).  37 galaxies at higher
latitudes were known from the optical survey.  28 of these could be
re-identified in \II, 26 in \J, and 14 in the \K\ band. In addition,
15 new galaxies were found in \II\ and \J, 11 of which also appear in
the \K\ band. The ratios of galaxies found in \II\ compared to \B, and
of \K\ compared to \II\, are higher than in the A3627 cluster. This is
due to the higher obscuration level (starting with $A_B \simeq 2\fm3
-3\fm1$ at the high-latitude border).

On average, about 3.5 galaxies per square degree were found in the
\II\ band. This roughly agrees with the predictions of
Fig.~\ref{nircts}. Because of star crowding, one does not expect to find
galaxies below latitudes of $b \simeq 1\deg$-$2\deg$ in this longitude
range \cite{Mam94}. 
Low-latitude images substantiate this -- the
images are nearly fully covered with stars.  Indeed, the lowest
Galactic latitude galaxies were found at $b \simeq 1\fdg2$ and $A_B
\simeq 11^{\rm m}$ (in \J\ and \K\ only).

\begin{figure} [thb]
\includegraphics[width=12cm]{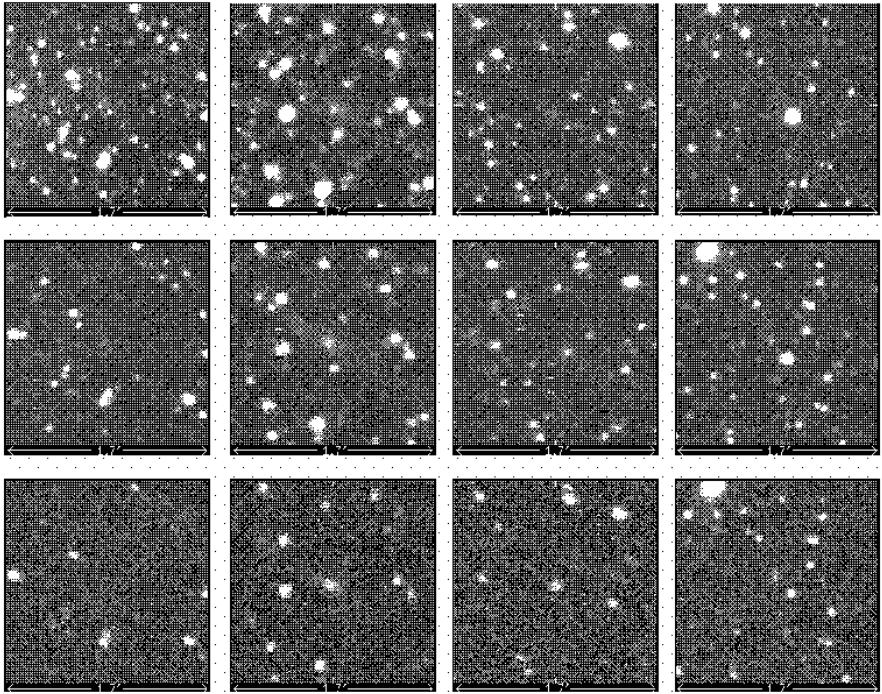}
\caption{DENIS survey images (before bad pixel filtering) of four
galaxies found in the deepest extinction layer of the Milky Way; 
the \II\  band image is at the top, \J\ in the middle and \K\ at the  bottom} 
\label{nirbsex}
\end{figure}

Figure~\ref{nirbsex} shows a few characteristic examples of highly
obscured galaxies found in the DENIS blind search. \II\ band images
are at the top, \J\ in the middle and \K\ at the bottom. The first
galaxy located at $(l,b) = (324\fdg6,-4\fdg5$) is viewed through an
extinction layer of $A_B = 2\fm0$ according to the DIRBE extinction
maps \cite{Sch98}. 
It is barely visible in the \J\ band. The
next galaxy at $(l,b) = (324\fdg7,-3\fdg5$) is subject to heavier
extinction ($A_B = 2\fm7$), and indeed easier to recognize in the
NIR. It is most distinct in the \J\ band. The third galaxy at even
higher extinction $(l,b,A_B) = (320\fdg1,+2\fdg5,5\fm7$) is -- in
agreement with the prediction of Fig.~\ref{nircts} -- not visible in
the \B\ band. Neither is the fourth galaxy at $b=+1\fdg9$ and $A_B =
9\fm6$: this galaxy can not be seen in \II\ band either and is very
faint only in \J\ and \K.

\subsection{Conclusions}

The conclusions from this pilot study are that at {\it intermediate
latitudes and extinction} ($|b| \GA 5\deg$, $A_B \LA 4$-$5^{\rm m}$)
optical surveys are superior for identifying galaxies. But despite the
extinction and the star crowding at these latitudes, \II , \J\ and \K\
photometry from the survey data could be performed successfully at
these low latitudes. The NIR data (magnitudes, colors) of these
galaxies can therefore add important data in the analysis of these
obscured galaxies.  They led, for instance, to the preliminary
$I_c^o$, $J^o$ and $K_s^o$ galaxy luminosity functions in A3627
(Fig.~2 in \cite{Kra98b}). 

At {\it lowest latitudes and high extinction} ($|b| \LA 5\deg$ and
$A_B \GA 4$-$5^{\rm m}$), the search for `invisible' obscured galaxies
on existing DENIS-images implicate that NIR-surveys can trace galaxies
down to about $|b| \GA 1\deg$-$1\fdg5$. The \J\ band was found to be
optimal for identifying galaxies up to $A_B \simeq 7^{\rm m}$.  NIR
surveys can hence further reduce the width of the ZOA.

The NIR surveys are particularly useful for the mapping of massive
early-type galaxies -- tracers of density peaks in the mass
distribution -- as these can not be detected with any of the
techniques that are efficient in tracing the spiral population 
in more opaque regions (Sect.~\ref{fir} and \ref{hi}).

Nevertheless, NIR surveys are also important with regard to the blue
and low surface-brightness spiral galaxies because a significant
fraction of them are also detectable in the near infrared. This is
confirmed, for instance, with the serendipitous discovery in the ZOA
of a large, nearby ($v = 750$~\kms) edge-on spiral galaxy by 2MASS
\cite{Hur99}: 
with an extension in the \K\ band of 5~arcmin, this
large galaxy is -- not unexpectedly for its extinction of $A_B =
6\fm6$ at the position of $(\ell,b) = (236\fdg8,-1\fdg8)$ -- not seen
in the optical \cite{Sai91}. 
Furthermore, the overlap of
galaxies found in NIR and \HI\ surveys allows the determination of
redshift independent distances via the NIR \tfr\ \cite{Tul77},
and therewith the peculiar velocity field.  This will provide
important new input on the mass density field ``in the ZOA''
(Sect.~\ref{recon}).

\section{Blind HI Surveys in the ZOA}\label{hi}

In the regions of the highest obscuration and infrared confusion, the
Galaxy is fully transparent to the 21cm line radiation of neutral
hydrogen. \HI-rich galaxies can readily be found at lowest latitudes
through the detection of their redshifted 21cm emission, though
early-type galaxies -- tracers of massive groups and clusters -- are
gas-poor and will not be identified in these surveys. Also very
low-velocity extragalactic sources might be missed due to the strong
Galactic \HI\ emission, and galaxies close to radio continuum sources.

An advantage of blind \HI\ surveys is the immediate availability of
rotational properties of a detected galaxy, next to its redshift,
providing insight on the intrinsic properties of these obscured
galaxies. The rotational velocity can furthermore be used (in
combination with \eg NIR photometry) to determine the distance in real
space from the \tfr, leading to determinations of the mass density
field from the peculiar velocities.

Until recently, radio receivers were not sensitive and efficient
enough to attempt systematic surveys of the ZOA. Kerr \& Henning
\cite{Ker87} 
demonstrated, however, the effectiveness of this approach: they
pointed the late 300-ft telescope of Green Bank to 1900 locations in
the ZOA (1.5\% coverage) and detected 19 previously unknown spiral
galaxies.

Since then two systematic blind \HI\ searches for galaxies behind the
Milky Way were initiated. The first -- the Dwingeloo Obscured Galaxies
Survey (DOGS) -- used the 25~m Dwingeloo radio to survey the whole
northern Galactic Plane for galaxies out to 4000~\kms\ 
\cite{Kra94b,Hen98,Riv99}.
A
more sensitive survey, probing a considerably larger volume (out to
12700 \kms), is being performed for the southern Milky Way at the 64~m
radiotelescope of Parkes \cite{Kra98a,Sta98,Hen99,Hen00}.

In the following, the observing techniques of these two surveys as
well as the first results will be discussed.

\subsection{The Dwingeloo Obscured Galaxies Survey}\label{dogs}

Since 1994, the Dwingeloo 25~m radio telescope has been dedicated to a
systematic search for galaxies in the northern Zone of Avoidance
($30\deg \le \ell \le 220\deg$, $|b| \le 5\fdg25$). The last few
patches of the survey were completed early 1999, using the Westerbork
array in total power mode.  The 20~MHz bandwidth was tuned to cover
the velocity range $0 \le v \le 4000$~\kms. 

The 25~m Dwingeloo telescope has a half-power-beamwidth (HPBW) of
36~arcmin. The 15000 survey points required for the survey coverage 
are ordered in a honeycomb pattern with a grid
spacing of $0\fdg4$. Galaxies are generally detected in various
adjacent pointings, facilitating a more accurate determination of
their positions through interpolations. The rms noise per channel typically
was $\sigma_{ch}=40$~mJy for a 1 hr integration (12 x 5min).

Because of the duration of the project (15000 hours not including
overhead and downtime) the strategy was to first conduct a fast search
of 5min integrations (${\rm rms}=175$~mJy) to uncover possible massive nearby
galaxies whose effect might yield important clues to the dynamics of
the Local Group. 

The shallow Dwingeloo search (${\rm rms}=175$~mJy) has been completed
in 1996 yielding five objects (\cf \cite{Hen98} 
for details),
three of which were known previously. The most exciting discovery was
the barred spiral galaxy Dwingeloo 1 \cite{Kra94b}. 

\begin{figure}[t]
\begin{center}
\includegraphics[width=12cm]{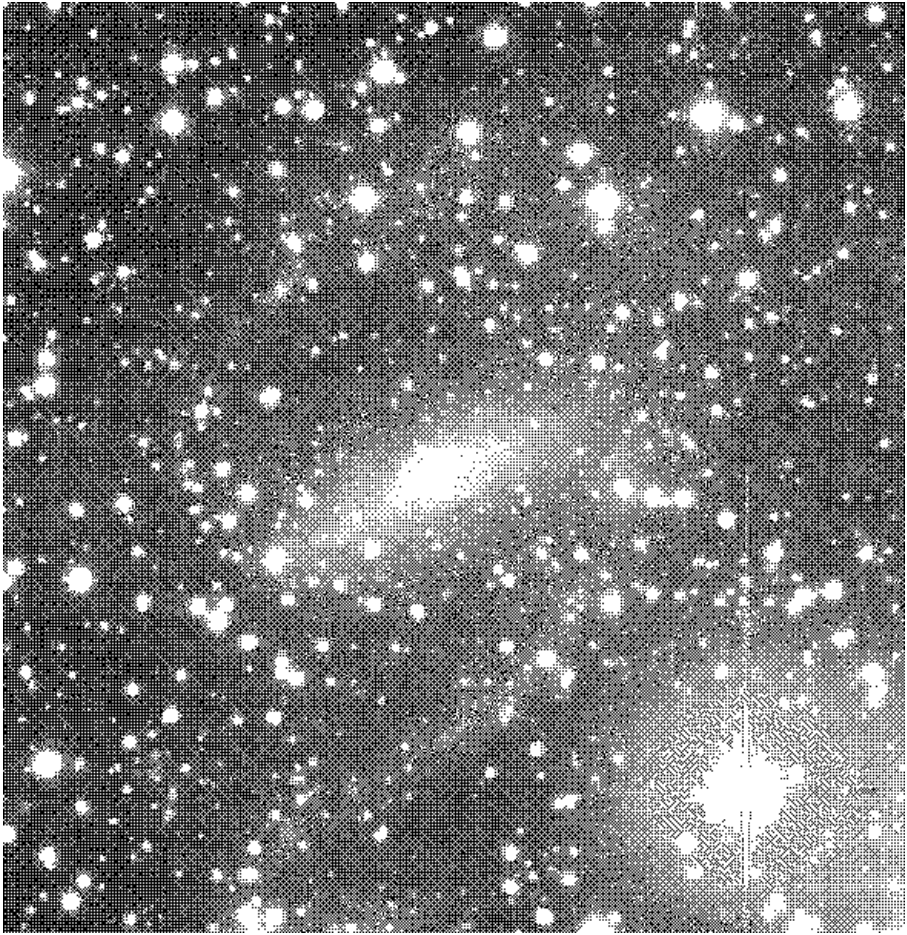}
\caption{Composite $V, R, I$-image of the Dwingeloo 1 galaxy at
$\ell=138\fdg5, b=-0\fdg1$. The displayed 484 x 484 pixels of
$0\farcs6$ cover an area of $4\farcm8$ x $4\farcm8$.  The large
diameter visible on this image is about $4\farcm2$.  Dwingeloo 1 has a
distinct bar, with 2 spiral arms that can be traced over nearly
$180\deg$. The morphology in this figure agrees with that of an SBb
galaxy
}
\label{Dw1}            
\end{center}
\end{figure}

This galaxy candidate was detected early on in the survey through a
strong signal (peak intensity of 1.4~Jy) at the very low redshift of
$v = 110$~\kms\ in the spectra of four neighboring
pointings, suggestive of a galaxy of large angular extent.  The
optimized position of $(\ell,b)=(138\fdg5,-0\fdg1)$ coincided
with a very low surface brightness feature on the Palomar Sky Survey
plate of $2\farcm2$, detected earlier by Hau \etal \cite{Hau95} 
in his
optical galaxy search of the northern Galactic/SuperGalactic Plane
crossing (\cf Sect.~\ref{optsear}). Despite foreground obscuration of
about 6$^{\rm m}$ in the optical, follow-up observations in the $V$,
$R$ and $I$ band at the INT (La Palma) confirmed this galaxy candidate
as a barred, possibly grand-design spiral galaxy of type SBb of 4.2 x
4.2 arcmin (\cf Fig.~\ref{Dw1}).

Dwingeloo 1 has been the subject of much follow-up observations
(optical: Loan \etal \cite{Loa96}, 
Buta \& McCall \cite{But99}; 
HI-synthesis: Burton\etal \cite{Bur96};
CO observations: Kuno \etal \cite{Kun96}, 
Li \etal \cite{Li96}, 
Tilanus \& Burton \cite{Til97}; 
X-ray: Reynolds \etal \cite{Rey97}). 
To summarize, it is a
massive barred spiral, with rotation velocity of 130~\kms, implying a
dynamical mass of roughly one-third the mass of the Milky Way.  Its
approximate distance of $\sim$ 3~Mpc and angular location place it
within the IC342/Maffei group of galaxies.  The follow-up HI synthesis
observations \cite{Bur96} 
furthermore revealed a
counterrotating dwarf companion, Dwingeloo 2. Since then various
further dwarf galaxies in this nearby galaxy group have been
discovered.

60\% of the deeper Dwingeloo survey (${\rm rms}=40$~mJy) has been analyzed
\cite{Riv99}. 
36 galaxies were detected, 23 of which were
previously unknown. Five of the 36 sources were originally identified
by the shallow survey. Based on the survey sensitivity, the registered
number of galaxies is in agreement with the Zwaan \etal \cite{Zwa97} 
HI mass
function which predicts 50 to 100 detections for the full survey.

Surprisingly, three dwarf galaxies were detected close to the nearby
isolated galaxy NGC 6946 at ($\ell,b,v) = (95\fdg7, 11\fdg7,
46$~\kms). One of these had earlier been catalogued as a compact High
Velocity Cloud \cite{Wak90}. 
Burton \etal \cite{Bur99}, 
in their search
for compact isolated high-velocity clouds in the Dwingeloo/Leiden
Galactic \HI\ survey \cite{Har94,Har97},
discovered a further member of this galaxy concentration.  Now, seven
galaxies with recessional velocities $v_{\rm_{LSR}} \le 250$~\kms\
have been identified within $15\deg$ of the galaxy NGC 6946. More
might be discovered as the DOGS data in this region have not yet been
fully analyzed.  The agglomeration of these various galaxies might
indicate a new group or cloud of galaxies in the nearby Universe.  As
such it would be the only galaxy group in the nearby Universe that is
strongly offset (by $40\deg$) from the Supergalactic Plane 
\cite{Tam78,Kra79}.

\subsection{The Parkes Multibeam ZOA Blind HI survey} \label{MB}

In March 1997, the systematic blind \HI\ survey in the southern Milky
Way ($212\deg \le \ell \le 36\deg$; $|b| \le 5\fdg5$) began with the
Multibeam receiver at the 64\,m Parkes telescope. The instrument has
13 beams each with a beamwidth of $14\farcm4$.  The beams are arranged
in a hexagonal grid in the focal plane array \cite{Sta96},
allowing rapid sampling of large areas.

The observations are being performed in driftscan mode. 23 contiguous
fields of length $\Delta\ell=8\deg$ have been defined.  Each field is
being surveyed along constant Galactic latitudes with latitude offsets
35~arcmin until the final width of $|b| \le 5\fdg5$ has been attained
(17 passages back and forth). The ultimate goal is 25 repetitions per
field. With an effective integration time of 25 min/beam a
3\,$\sigma$ detection limit of 25\,mJy is obtained.  The survey covers
the velocity range $-1200 \LA v \LA 12700$~\kms\ and will be sensitive
to normal spiral galaxies well beyond the Great Attractor region.

So far, a shallow survey covering the whole southern Milky Way based
on 2 out of the foreseen 25 driftscan passages has been analyzed (\cf
\cite{Kra98a,Hen99,Hen00}).
A detailed
study of the Great Attractor region ($308\deg \le \ell \le 332\deg$)
based on 4 scans has been made by Juraszek \etal \cite{Jur99,Jur00}.
The first four full-sensitivity cubes are available for that
region as well (Sect.~\ref{MBdeep}).

In the shallow survey, 110 galaxies were catalogued with peak \HI-flux
densities of $\GA$80~mJy (${\rm rms} = 15$~mJy after Hanning
smoothing). The detections show no dependence on Galactic latitude,
nor the amount of foreground obscuration through which they have been
detected.  Though galaxies up to 6500~\kms\ were identified, most of
the detected galaxies (80\%) are quite local ($v<3500$\kms) due to the
(yet) low sensitivity. About one third of the detected galaxies have a
counterpart either in NED (NASA/IPAC Extragalactic Database) or in the
deep optical surveys.

\begin{figure}[ht]
\begin{center}
\includegraphics[width=12cm]{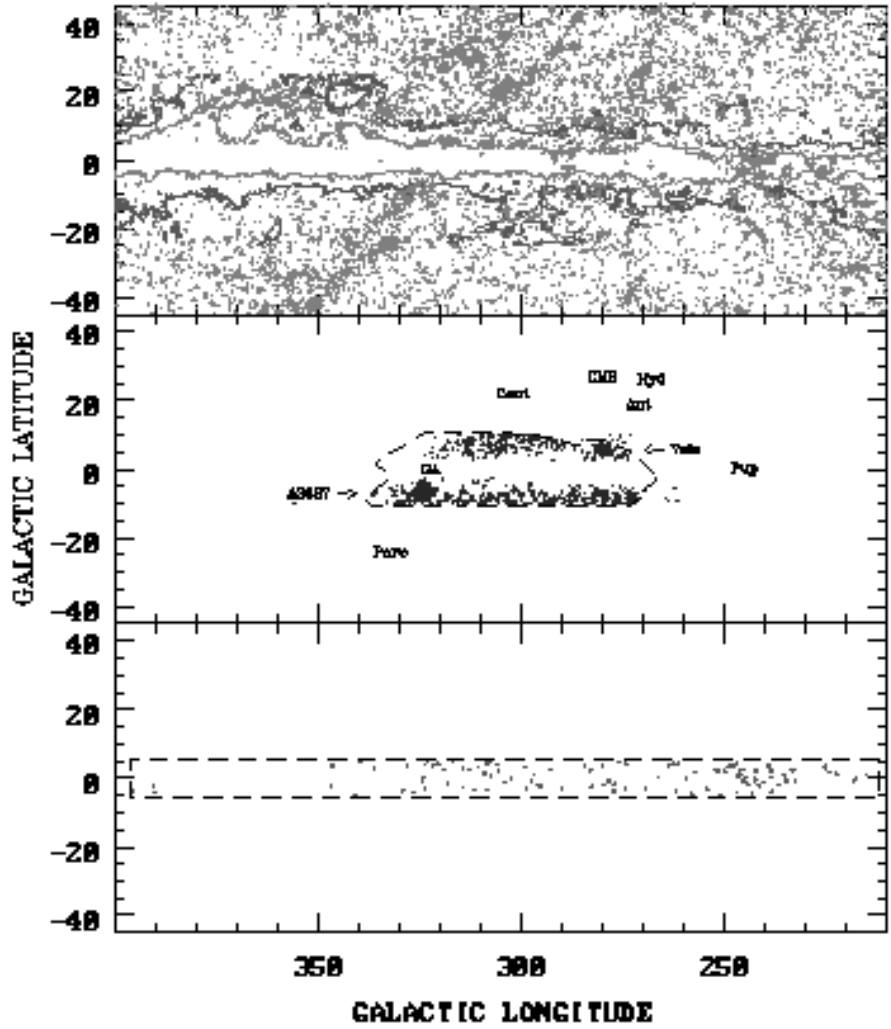}
\caption{Galaxies with $v<10000$~\kms.
Top panel: literature values (LEDA), superimposed
are extinction levels A$_B=1\fm0$ and $3\fm0$; middle panel:
follow-up redshifts (ESO, SAAO and Parkes) from deep optical
ZOA survey with locations of clusters and dynamically important
structures; bottom panel: galaxies detected with the shallow Multibeam
ZOA survey}
\label{vall}
\end{center}
\end{figure}

The distribution of the 110 \HI-detected galaxies is displayed in the
lower panel of Fig.~\ref{vall}. It demonstrates convincingly that
galaxies can be traced through the thickest extinction layers of the
Galactic Plane. The fact that hardly any galaxies are found behind the
Galactic bulge ($\ell=350\deg$ to $\ell=30\deg$) is due to local
structure: this is the region of the Local Void.

For comparative purposes, the top panel of Fig.~\ref{vall} shows the
distribution of all known galaxies with $v\le 10000$~\kms\ (extracted
from the Lyon-Meudon Extragalactic Database (LEDA). Although this
constitutes an uncontrolled sample, it traces the main structures in
the nearby Universe in a representative way.  Note the increasing
incompleteness for extinction levels of ${A}_B \GA 1\fm0$ (outer
contour) -- reflecting the growing incompleteness of optical galaxy
catalogs -- and the near full lack of galaxy data for extinction
levels ${A}_B \GA 3\fm0$ (inner contour).  The middle panel shows
galaxies with $v<$10000 \kms\ from the follow-up observations of the
deep optical galaxy search by Kraan-Korteweg and collaborators
(Sect.~\ref{optred}). Various new overdensities are apparent at low
latitudes but the innermost part of our Galaxy remains obscured with
this approach. Here, the blind \HI\ data (lower panel) finally can
provide the missing link for large-scale structure studies.

\begin{figure}[ht]
\begin{center}
\includegraphics[width=12cm]{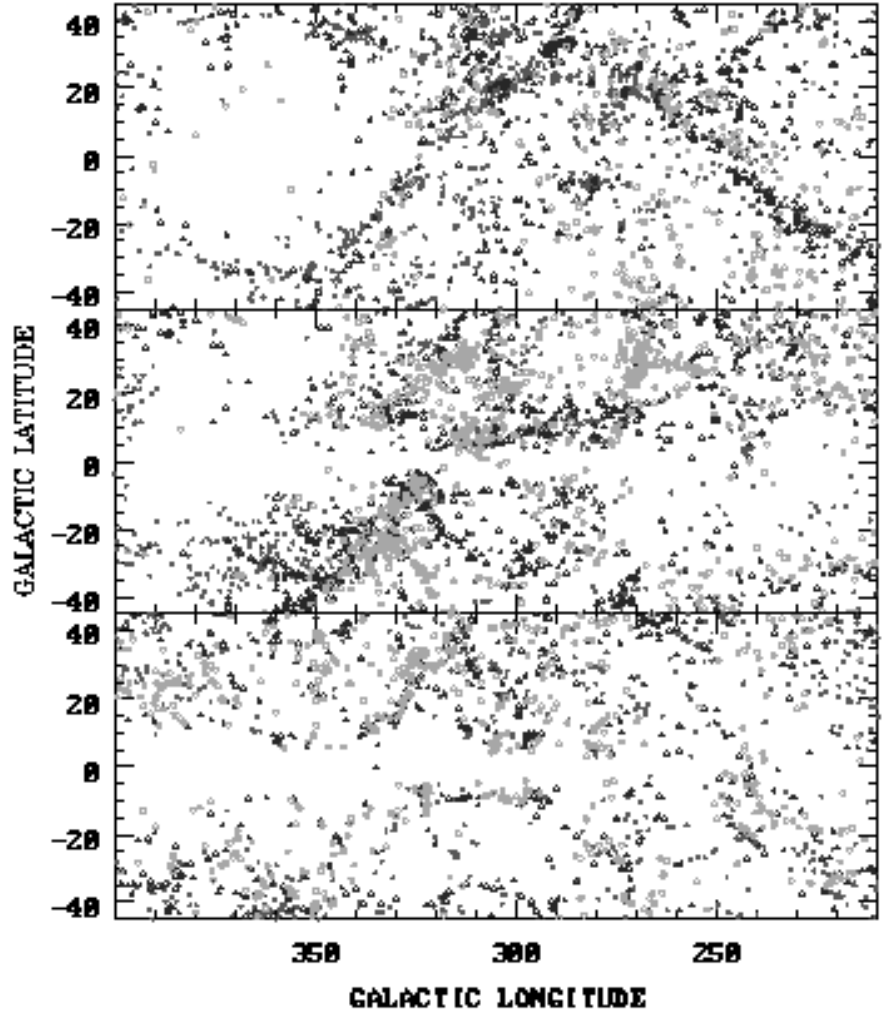}
\caption{Redshift slices from the data in Fig.~\ref{vall}: $500<v<3500$
(top), $3500<v<6500$ (middle), $6500<v<9500$~\kms\
(bottom). The open circles mark the nearest $\Delta v=1000$~\kms\ slice
in a panel, then triangles, then the filled dots the 2 more distant ones}
\label{vslice}
\end{center}
\end{figure}

In Fig.~\ref{vslice}, the data of Fig.~\ref{vall} are combined in
redshift slices. The achieved sensitivity of the shallow MB \HI-survey
fills in structures all the way across the ZOA for $v<$ 3500~\kms
(upper panel) for the first time. Note the continuity of the thin
filamentary sine-wave-like structure that dominates the whole southern
sky and crosses the Galactic Equator twice. This structure snakes over
$\sim180\deg$ through the southern sky. Taking a mean distance of
$30{h^{-1}}$ Mpc, this implies a linear size of $\sim100{h^{-1}}$ Mpc,
with a thickness of 'only' $\sim5{h^{-1}}$ Mpc or less. Various other
filaments spring forth from this dominant filament, always from a rich
group or small cluster at the junction of these interleaving
structures. This feature is very different from the thick, foamy Great
Wall-like structure, the GA, in the middle panel.

Also note the prominence of the Local Void which is very well
delineated in this presentation. No galaxies were found
within the Local Void, but the three newly identified galaxies at
$\ell \sim 30\deg$ help to define the boundary of the Void.

The full sensitivity ZOA MB-survey will fill in the large-scale
structures in the more distant panels of Fig.~\ref{vslice}.
First results of the full sensitivity survey have been obtained in the
Great Attractor region (Sect.~\ref{MBdeep}).

Three nearby, very extended ($20\arcmin$ to $\GA 1\deg$) galaxies were
discovered with the shallow survey. Being likely candidates of
dynamically important galaxies, immediate follow-up observations were
initiated at the Australian Telscope Compact Array (ATCA). These
objects did not turn out to be massive perturbing monsters, however.
Two were seen to break up into \HI\ complexes and both have
unprecedented low \HI\ column densities \cite{Sta98}.
Systematic synthesis observations are being performed to
investigate the frequency of these interacting and/or low \HI\ column
density systems in this purely \HI-selected sample.

\subsection{The Parkes ZOA MB Deep Survey and the Great Attractor} 
\label{MBdeep}

Four cubes centered on the Great Attractor region ($300\deg \ge \ell
\ge 332\deg$, $|b| \le 5\fdg5$) of the full-sensitivity survey have
been analyzed \cite{Jur00}. 
236 galaxies above the $3\sigma$ 
detection level of 25~mJy have been uncovered. 70\% of the detections
had no previous identification.

\begin{figure}[ht]
\begin{center}
\includegraphics[width=12cm]{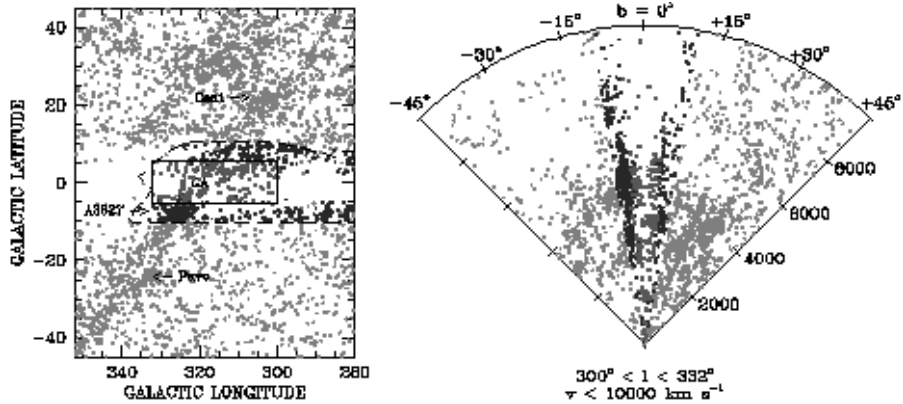}
\caption
{A sky distribution (left) and redshift cone (right) for galaxies with
$v<10000$~\kms\ in the GA region. Circles mark redshifts from the
literature (LEDA), squares redshifts from the optical galaxy search in
the Hy/Ant-Crux-GA regions (outlined on left panel) and crosses
detections in the full-sensitivity HI MB-ZOA survey (box)}
\label{MBGA}
\end{center}
\end{figure}

In the left panel of Fig.~\ref{MBGA}, a sky distribution centered on
the GA region displays all galaxies with redshifts ${v} \le
10000$~\kms. Next to redshifts from the literature, redshifts from the
follow-up observations of Kraan-Korteweg and collaborators in the
Hy/Ant-Crux-GA ZOA surveys (dashed area) are plotted. They clearly
reveal the prominence of the cluster A3627 at $(\ell,b,v) =
(325\deg, -7\deg, 4882$~\kms) close to the core of the GA region at
$(\ell,b,v)$ = (320$\deg$, 0$\deg$, 4500~\kms).  Adding now the new
detections from the systematic blind \HI\ MB-ZOA survey (box),
structures can be traced all the way across the Milky Way. The new
picture seems to support that the GA overdensity is a ``great-wall''
like structure starting close to the Pavo cluster, having its core at
the A3627 cluster and then bending over towards shorter longitudes
across the ZOA.

This becomes even clearer in the right panel of Fig.~\ref{MBGA} 
(compare with right hand panel of Fig.~\ref{pie}) where
the galaxies are displayed in a redshift cone out to ${v} \le
10000$~\kms\ for the longitude range $300\deg \le \ell \le 332\deg$.
The combined surveys in the GA region clearly substantiate that A3627
is the most massive galaxy cluster uncovered in this region and
therefore the most likely candidate for the predicted density-peak at
the bottom of the potential well of the GA overdensity.  The new data
do not unambigously confirm the existence of the suspected further
cluster around the bright elliptical radio galaxy PKS1343$-$601
(Sect.~\ref{optred}). Although the MB data reveal an excess of
galaxies at this position in velocity space ($b=+2\deg, v=4000$~\kms)
a ``finger of God'' is not seen. It could be that many central cluster
galaxies are missed by the \HI\ observations because spiral galaxies
generally avoid the cores of clusters. The reality of this possible
cluster still remains a mystery. This prospective cluster has
meanwhile been imaged in the $I$-band \cite{Wou00c}, 
where
extinction effects are less severe compared to the optical (see
Sect.~\ref{nir}). A first glimpse of the images do reveal various
early-type galaxies.  The forthcoming analysis should then
unambiguously settle the question whether another cluster forms part
of the GA overdensity.

\subsection{Conclusions}

The systematic probing of the galaxy distribution in the most opaque
parts of the ZOA with \HI\ surveys have proven very powerful. For the
first time large-scale structure could be mapped without hindrance
across the Milky Way (Figs.~\ref{vslice} and \ref{MBGA}). This is the
only approach that easily uncovers the galaxy distribution in the ZOA,
allows the confirmation of implied connections and uncovers new
connections behind the Milky Way.

>From the analysis of the Dwingeloo survey and the shallow Parkes MB
ZOA survey, it can be maintained that no Andromeda or other \HI-rich
Circinus-like galaxy is lurking undetected behind the deepest
extinction layers of the Milky Way (although gas-poor, early-type
galaxies might, of course, still remain hidden). The census of
dynamically important, \HI-rich nearby galaxies whose gravitational
influence could significantly impact peculiar motion of the Local
Group or its internal dynamics is now complete -- at least for objects
whose signal is not drowned within the strong Galactic \HI\ emission.

\section{X-ray Surveys} \label{Xray}

The X-ray band potentially is an excellent window for studies of
large-scale structure in the Zone of Avoidance, because the Milky Way
is transparent to the hard X-ray emission above a few keV, and because
rich clusters are strong X-ray emitters. Since the X-ray luminosity is
roughly proportional to the cluster mass as $L_X \propto M^{3/2}$ or
$M^2$, depending on the still uncertain scaling law between the X-ray
luminosity and temperature, massive clusters hidden by the Milky Way
should be easily detectable through their X-ray emission.

This method is particularly attractive, because clusters are primarily
composed of early-type galaxies which are not recovered by IRAS galaxy
surveys (Sect.~\ref{fir}) or by systematic \HI\ surveys
(Sect.~\ref{hi}). Even in the NIR, the identification of early-type
galaxies becomes difficult or impossible at the lowest Galactic
latitudes because of the increasing extinction and crowding problems
(Sect.~\ref{nir}).  Rich clusters, however, play an important role in
tracing large-scale structures because they generally are located at the
center of superclusters and Great Wall-like structures. They mark the
density peaks in the galaxy distribution and -- with the very high
mass-to-light ratios of clusters -- the deepest potential wells within these
structures.  Their location within these overdensities will help us
understand the observed velocity flow fields induced by these
overdensities.

The X-ray all-sky surveys carried out by Uhuru, Ariel V, HEAO-1 (in
the $2$-$10$~keV band) and ROSAT ($0.1$-$2.4$~keV) provide an optimal
tool to search for clusters of galaxies at low Galactic
latitude. However, confusion with Galactic sources such as X-ray
binaries and Cataclysmic Variables may cause serious problems,
especially in the earlier surveys Uhuru, Ariel V and HEAO-1 which had
quite low angular resolution.  And although dust extinction and
stellar confusion are unimportant in the X-ray band, photoelectric
absorption by the Galactic hydrogen atoms -- the X-ray absorbing
equivalent hydrogen column density -- does limit detections close to
the Galactic Plane. The latter effect is particularly severe for the
softest X-ray emission, as \eg observed by ROSAT ($0.1$-$2.4$~keV)
compared to the earlier $2$-$10$~keV missions. On the other hand, the
better resolution of the ROSAT All Sky Survey (RASS), compared to the
HEAO-1 survey, will reduce confusion problems with Galactic sources as
happened, for example, in the case of the cluster A3627 (see below).

Until recently, the possibility of searching for galaxy
clusters behind the Milky Way through their X-ray emission has not 
been pursued in a systematic way, even though a large number of 
X-ray bright clusters are located at low Galactic latitudes \cite{Fab94}:
for instance, four of the seven most X-ray luminous
clusters in the 2-10 keV range, the Perseus, Ophiuchus,
Triangulum Australis, and PKS0745$-$191  clusters ($L_{\rm X}>10^{45}$ 
erg s$^{-1}$) lie at latitudes below $|b|<20^{\circ}$ \cite{Edg90}.

A first attempt to identify galaxy clusters in the ZOA through their
X-ray emission had been made by Jahoda and Mushotzky in 1989 \cite{Jah89}. 
They used the HEAO-1 all-sky data to search for X-ray-emission of a
concentration of clusters or one enormous cluster that might help
explain the shortly before discovered large-scale deviations from the
Hubble flow that were associated with the Great Attractor.
Unfortunately, this search missed the 6$^{th}$ brightest cluster A3627
in the ROSAT X-ray All Sky Survey \cite{Boh96,Tam98}
which had been identified as the most likely candidate for
the predicted but unidentified core of the Great Attractor.  A3627 was
not seen in the HEAO-1 data because of the low angular resolution and
the confusion with the neighbouring X-ray bright, Galactic X-ray
binary 1H1556-605 (\cf Fig.~8 and 9 in \cite{Boh96}). 

\subsection{CIZA: Clusters In the Zone of Avoidance}

Since 1997, a group led by Ebeling \cite{Ebe99a,Ebe99b} 
have
systematically searched for bright X-ray clusters of galaxies at $|b|
< 20\deg$. Starting from the ROSAT Bright Source Catalog (BSC, 
\cite{Vog99}) 
which lists the 18811 X-ray brightest sources detected in
the RASS, they apply the following criteria to search for clusters:
(a) $|b|<20^{\circ}$, (b) a X-ray flux above $S > 5\times 10^{-12}$
erg cm$^{-2}$ s$^{-1}$ (the flux limit of completeness of the ROSAT
BCS), and (c) a spectral hardness ratio. Ebeling \etal demonstrated in
1998 that the X-ray hardness ratio is very effective in discriminating
against softer, non-cluster X-ray sources. With these criteria, they
select a candidate cluster sample which, although at this point still
highly contaminated by non-cluster sources, contains the final CIZA
cluster sample.

They first cross-identified their 520 cluster candidates against NED
and SIMBAD, and checked unknown ones on the Digitized Sky Survey.  The
new cluster candidates, including known Abell clusters without
photometric and spectroscopic data, were imaged in the R band,
respectively in the K' band at high extinctions. With the subsequent
spectroscopy of galaxies around the X-ray position, the real clusters
could be confirmed.

Time and funding permitting, the CIZA team plans to extend their
cluster survey to lower X-ray fluxes ($2$-$3\times 10^{-12}$ erg 
cm$^{-2}$ s$^{-1}$), the aim being a total sample of 200 X-ray selected
clusters below $|b| < 20\deg$.

\begin{figure}[ht]
\begin{center}
\includegraphics[width=12cm]{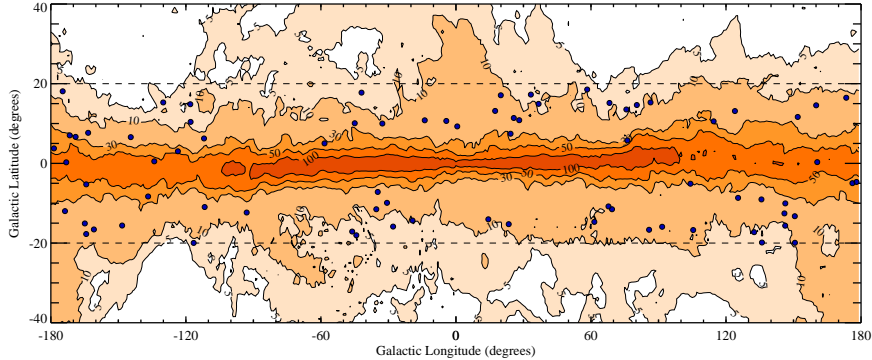}
\caption
{Distribution in Galactic coordinates of the 76 by Ebeling \etal [129]
so far spectroscopically confirmed X-ray clusters (solid dots)
of which 80\% were previously unknown. Superimposed are Galactic HI
column densities in units of $10^{20}$ cm$^{-2}$ (Dickey \& Lockman
1990).  Note that the region of relatively high absorption ($N_{\rm HI}
> 5 \times 10^{21}$ cm$^{-2}$) actually is very narrow and that
clusters could be identified to very low latitudes
}
\label{xray}
\end{center}
\end{figure}

So far, 76 galaxy clusters were identified within $|b| < 20\deg$ of
which 80\% were not known before. Their distribution (reproduced
from Ebeling \etal \cite{Ebe99b}) 
is displayed in Fig.~\ref{xray}. 14 of
these clusters are relatively nearby ($z \le 0.04$), and one was
uncovered at a latitude of only $b = 0\fdg3$ within the Perseus-Pisces
chain.

\subsection{Conclusions}

With the discovery of so far 76 clusters of which only 20\% were known
before, Ebeling \etal \cite{Ebe99b} 
have proven the strength of the method to
use X-ray criteria to search for galaxy clusters in the ZOA. As
mentioned in the introduction to this section, this approach is 
complementary to the other wavelengths searches which all fail to
uncover galaxy clusters at very low Galactic latitudes. 

Having used the ROSAT BSC to select their galaxy cluster candidates,
the CIZA collaboration can combine their final cluster sample with
other X-ray selected cluster samples from the RASS, such as the ROSAT
Brightest Cluster Sample at $|b|\ge 20^{\circ}$ and $\delta \ge 0\deg$
\cite{Ebe98} 
and the REFLEX sample at $|b|\ge 20\deg$
and $\delta \le 2.5\deg$ (B\"ohringer \etal in prep.). The resulting,
all-sky cluster list will be ideally suited to study large-scale
structure and the connectivity of superclusters across the Galactic Plane.

\section{Theoretical Reconstructions} \label{recon}

Various mathematical methods exist to reconstruct the galaxy 
distribution in the ZOA without having access to direct observations.

One possibility is the expansion of galaxy distributions
adjacent to the ZOA into spherical harmonics to recover
the structures in the ZOA, either with 2-dimensional catalogs (sky
positions)  or 3-dimensional data sets (redshift catalogs). 

A statistical method to reconstruct structures behind the Milky Way is
the Wiener Filter (WF), developed explicitly for reconstructions of
corrupt or incomplete data \cite{Lah94,Hof94}.
Using the 
WF in combination with linear theory allows the determination
of the real-space density of galaxies, as well as their velocity
and potential fields.

The POTENT analysis developed by \cite{Ber89} 
can reconstruct the potential field (mass distribution) from
peculiar velocity fields in the ZOA \cite{Kol95}. 
The  reconstruction of the potential fields versus density fields 
have the advantage that they can locate hidden overdensities  (their 
signature) even if ``unseen''. 

Because of the sparsity of data and
the heavy smoothing applied in all these methods, only structures 
on large scales (superclusters) can be mapped.
Individual (massive) nearby galaxies that can perturb the
dynamics of the Universe quite locally (the vicinity of the Local
Group or its barycenter) will not be uncovered in this manner. 
But even if theoretical methods can outline LSS
accurately, the observational efforts do not become
superfluous. The comparison of the real galaxy distribution 
$\delta_{g}$ ({\bf r}), from \eg\ complete redshift surveys, with the
peculiar velocity field {\bf v}({\bf r}) will lead to an estimate of the 
density and biasing parameter ($\Omega^{0.6}/{b}$)  through the
equation 
\begin{equation}
 \nabla \cdot {\bf v (r)} = - {{\Omega^{0.6}}\over{b}} \, \, \delta_g
({\bf r}),
\end{equation}
\cf Strauss \& Willick \cite{Str95} 
for a detailed review.

\subsection{Early Predictions}

Early reconstructions on relatively sparse data galaxy catalogs have
been performed within volumes out to $v \le$ 5000~\kms. Despite
heavy smoothing, they have been quite successful in 
pinpointing a number of important features: 

$\bullet$ Scharf \etal \cite{Sch92} 
applied spherical harmonics to the
2-dimensional IRAS PSC and noted a prominent cluster behind the ZOA in
Puppis ($\ell \sim 245\deg$) which was simultaneously discovered as a
nearby cluster through \HI-observations of obscured galaxies in that
region by Kraan-Korteweg \& Huchtmeier \cite{Kra92}. 

$\bullet$ Hoffman \cite{Hof94} 
predicted the Vela supercluster at ($280\deg,
6\deg, 6000$~\kms) using 3-dimensional WF reconstructions on the IRAS 1.9
Jy redshift catalog \cite{Str92}, 
which was observationally
discovered just a bit earlier by Kraan-Korteweg \& Woudt \cite{Kra93}. 

\begin{figure}[ht]
\begin{center}
\includegraphics[angle=-90,width=12cm]{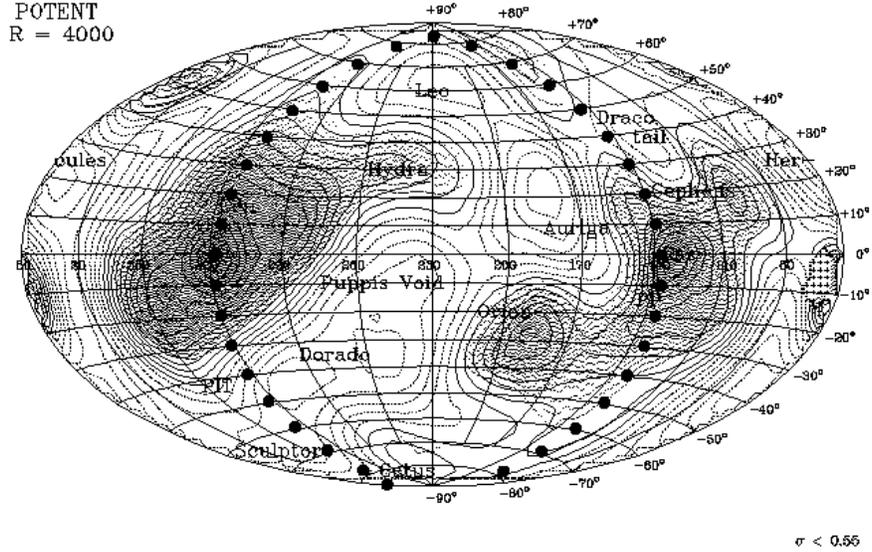}
\caption
{The mass-density fluctuation field in a shell at 4000~\kms\
as determined with POTENT from peculiar velocity data. The 
density is smoothed by a three-dimensional Gaussian of radius 
1200~\kms. Density contour spacings are $\Delta\delta = 0.1$ with 
$\delta=0$ as a heavy contour. Compared to Fig.~\ref{ait} and
\ref{aitc} this Aitoff projection is displaced by $\Delta \ell =
50\deg$. The Supergalactic Plane is indicated (solid dots).
(Figure 1b from [19])}
\label{kolatt}
\end{center}
\end{figure}

$\bullet$ Using POTENT analysis, Kolatt \etal \cite{Kol95} 
predicted the
center of the Great Attractor overdensity -- its density peak -- to
lie behind the ZOA at ($320\deg, 0\deg, 4500$~\kms, see
Fig.~\ref{kolatt}). Shortly thereafter, Kraan-Korteweg \etal \cite{Kra96}
unveiled the cluster A3627 as being very rich and massive and at the
correct distance. It hence is the most likely candidate for the
central density peak of the GA.

\subsection{Deeper Reconstructions}

Recent reconstructions have been applied to denser galaxy samples
covering larger volumes ($v \LA 10000$~\kms) with smoothing scales of
the order of 500~\kms\ (compared to 1200~\kms in the earlier
reconstructions). It therefore seemed of interest to see whether these
reconstructions find evidence for unknown major galaxy structures at
higher redshifts.

The currently most densely-sampled, well-defined galaxy redshift
catalog is the Optical Redshift Survey \cite{San95}. 
However, this catalog is limited to $|b| \ge 20\deg$ and the
reconstructions \cite{Bak98} 
within the ZOA are strongly
influenced by 1.2~Jy IRAS Redshift Survey data and a mock galaxy
distribution in the inner ZOA.  I therefore concentrate on
reconstructions based on the 1.2~Jy IRAS Redshift Survey only. In the
following, the structures identified in the ZOA by (a) Webster \etal
\cite{Web97} 
using WF plus spherical harmonics and linear theory and (b)
Bistolas \cite{Bis98} 
who applied a WF plus linear theory and
non-constrained realizations on the 1.2~Jy IRAS Redshift Survey are
discussed and compared to observational data.  Fig.~2 in Webster
\etal displays the reconstructed density fields on shells of 2000,
4000, 6000 and 8000~\kms; Fig.~5.2 in Bistolas displays the density
fields in the ZOA from 1500 to 8000~\kms\ in steps of 500~\kms.

The WLF reconstructions clearly find the recently by Roman \etal 
\cite{Rom98} 
identified nearby cluster at ($33\deg$, $5\deg$-$15\deg$,
1500~\kms), whereas Bistolas reveals no clustering in the region of
the Local Void out to 4000~\kms. At the same longitudes, clustering is
indicated at 7500~\kms\ by Bistolas, but not by Webster \etal\ The
Perseus-Pisces chain is strong in both reconstructions, and the 2nd
Perseus-Pisces arm -- which folds back at $\ell\sim 195\deg$ -- is
clearly confirmed.  Both reconstructions find the Perseus-Pisces
complex to be very extended in space, \ie\ from 3500~\kms\ out to
9000~\kms.  Whereas the GA region is more prominent compared to
Perseus-Pisces in the Webster \etal reconstructions, the signal of the
Perseus-Pisces complex is considerably stronger than the GA in
Bistolas, where it does not even reveal a well-defined central density
peak. Both reconstructions find no evidence for the suspected PKS1343
cluster but its signal could be hidden in the central (A3627) density
peak due to the smoothing.  While the Cygnus-Lyra complex
($60\deg$-$90\deg, 0\deg, 4000$~\kms) discovered by Takata \etal \cite{Tak96}
stands out clearly in Bistolas, it is not evident in Webster \etal
Both reconstructions find a strong signal for the Vela supercluster ($285\deg,
6\deg, 6000$~\kms) identified by Kraan-Korteweg \& Woudt \cite{Kra93}
and Hoffman \cite{Hof94}. 
The Cen-Crux cluster identified by Woudt \cite{Wou98}
is evident in Bistolas though less distinct in Webster \etal\ A suspected
connection at ($\ell,v) \sim (345\deg, 6000$~\kms) -- \cf Fig.~2 in
\cite{Kra98a} 
-- is supported by both methods.  The
Ophiuchus cluster \cite{Has00} 
just becomes visible in the
most distant reconstruction shells (8000~\kms).

\subsection{Conclusions}

Not all reconstructions find the same features, and when they do, the
prominence of the density peaks as well as their locations in space do
vary considerably. At velocities of $\sim 4000$~\kms\ most of the
dominant structures lie close to the ZOA while at larger distances,
clusters and voids seem to be more homogeneously distributed over the
whole sky. Out to 8000~\kms, none of the reconstructions predict any
major structures which are not mapped or suggested from observational
data. So, no major surprises seem to remain hidden in the ZOA. The
various multi-wavelength explorations of the Milky Way will soon be
able to verify this. Still, the combination of both the reconstructed
potential fields and the observationally mapped galaxy distribution
will lead to estimates of the cosmological parameters $\Omega_0$ and
$b$.

\section{Conclusions}

In the last decade, enormous progress has been made in unveiling the
extragalactic sky behind the Milky Way. At optical wavebands, the
entire ZOA has been systematically surveyed. It has been shown that
these surveys are complete for galaxies larger than $D^o = 1\farcm3$
(corrected for absorption) down to extinction levels of ${A_B} =
3\fm0$. Combining these data with previous ``whole-sky'' maps results
in a reduction of the ``optical ZOA'' of a factor of about 2-2.5
which allow an improved understanding of the velocity flow fields and
the total gravitational attraction on the Local Group. Various
previously unknown structures in the nearby Universe could be mapped
in this way.

At higher extinction levels, other windows to the ZOA become more
efficient in tracing the large-scale structures.  Very promising in
this respect are the current near-infrared surveys which find galaxies
down to latitudes of $|b| \sim 1\fdg5$ and systematic \HI\ surveys
which detect gas-rich spiral galaxies all the way across the Galactic
Plane -- hampered slightly only at very low latitudes ($|b|\LA
1\fdg0$) because of the numerous continuum sources.  The ``Behind the
Plane'' Survey resulted in a reduction from 16\% to 7\% of the ``FIR
ZOA'' and new indications of possible hidden massive clusters behind
the Milky Way are now forthcoming from the CIZA project -- although
again an ``X-ray ZOA'' will remain due to the absorption of X-ray
radiation by the thickening gas layer close to the Galactic Plane.

A difficult task is still awaiting us, \ie to obtain a detailed
understanding of the selection effects inherent to the various 
methods in order to merge the different data sets in a uniform,
well-defined way. This is extremely important if we want to use this
data for quantitative cosmography. Moreover, we need a better
understanding of the obscurational effects on the observed properties
of galaxies identified through the dust layer (at all wavelengths), in
addition to an accurate high-resolution, well-calibrated map of the
Galactic extinction.

Despite the fact that our knowledge of the above questions is as yet
limited, a lot can and has been learned from ZOA research. This is
evident, for instance, from the detailed and varied investigations of
the Great Attractor region. Mapping the GA and understanding the from
peculiar velocity fields inferred massive overdensity had remained an
enigma due the fact that the major and central part of this extended
density enhancement is largely hidden by the obscuring veil of the
Milky Way. Does light trace mass in this region and where is the
rich cluster which biasing predicts at the center of large-scale
potential wells?

The results from the various ZOA surveys now clearly imply that the
Great Attractor is, in fact, a nearby ``great-wall'' like
supercluster, starting at the nearby Pavo cluster below the GP, moving
across the massive galaxy cluster A3627 toward the shallow overdensity
in Vela at 6000~\kms. The cluster A3627 is the dominant central
component of this structure, similar to the Coma cluster in the
(northern) Great Wall. Whether a second massive cluster around
PKS1343$-$601 is part of the core of the GA remains uncertain.

\vspace{0.5cm}

\noindent {\bf Acknowledgement.}
The enthusiastic collaborations of my colleagues in the exploration of the
galaxy distribution behind the Milky Way is greatly appreciated. These
are P.A. Woudt, C. Salem and A.P. Fairall with deep optical searches,
C. Balkowski, V. Cayatte, A.P. Fairall, P.A. Henning with the redshift
follow-ups of the optically identified galaxies, A. Schr\"oder and
G.A. Mamon in the exploration of the DENIS images at low Galactic
latitude, W.B. Burton, P.A. Henning, O. Lahav and A. Rivers in the
northern ZOA HI-survey (DOGS) and the HIPASS ZOA team members
L. Staveley-Smith , R.D. Ekers, A.J. Green, R.F. Haynes, P.A. Henning,
S. Juraszek, M. J. Kesteven, B. Koribalski, R.M. Price, E. Sadler and
A. Schr\"oder in the southern ZOA survey.

Particular thanks go to P.A. Woudt for his careful reading of the
manuscript and his valuable suggestions, to W. Saunders for preparing
Fig.~\ref{BTP}, to A. Schr\"oder and G. Mamon for their comments on the
NIR section, and to H. Ebeling for his input with regard
to the X-ray section and Fig.~\ref{xray}.

This research has made use of the NASA/IPAC Extragalactic Database (NED)
which is operated by the Jet Propulsion Laboratory, Caltech, under
contract with the National Aeronautics and Space Administration,
as well as the Lyon-Meudon Extragalactic Database (LEDA),
supplied by the LEDA team at the Centre de Recherche Astronomique de
Lyon, Observatoire de Lyon.

%
\clearpage
\addcontentsline{toc}{section}{Index}
\flushbottom
\printindex

\end{document}